\documentclass[
reprint,
 amsmath,amssymb,
 aps,
]{revtex4-2}

\usepackage{graphicx}% Include figure files
\usepackage{dcolumn}% Align table columns on decimal point
\usepackage{bm}% bold math
\usepackage{hyperref}% add hypertext capabilities
\usepackage{physics}

\begin{document}

\title{High-rate and high-fidelity modular interconnects between neutral atom quantum processors}% Force line breaks with \\

\author{Yiyi Li}%
 \affiliation{Department of Electrical and Computer Engineering, Princeton University, Princeton, New Jersey 08544, USA}

\author{Jeff D. Thompson}%
 \affiliation{Department of Electrical and Computer Engineering, Princeton University, Princeton, New Jersey 08544, USA}

\date{\today}

\begin{abstract}
Quantum links between physically separated modules are important for scaling many quantum computing technologies. The key metrics are the generation rate and fidelity of remote Bell pairs. In this work, we propose an experimental protocol for generating remote entanglement between neutral ytterbium atom qubits using an optical cavity. By loading a large number of atoms into a single cavity, and controlling their coupling using only local light shifts, we amortize the cost of transporting and initializing atoms over many entanglement attempts, maximizing the entanglement generation rate. A twisted ring cavity geometry suppresses many sources of error, allowing high fidelity entanglement generation. We estimate a spin-photon entanglement rate of $5 \times 10^5$~s$^{-1}$, and a Bell pair rate of $1.0\times 10^5$~s$^{-1}$, with an average fidelity near $0.999$. Furthermore, we show that the photon detection times provide a significant amount of soft information about the location of errors, which may be used to improve the logical qubit performance. This approach provides a practical path to scalable modular quantum computing using neutral ytterbium atoms.
\end{abstract}

\maketitle

\section{Introduction}

The development of a large-scale, fault-tolerant quantum computer capable of solving classically intractable problems is expected to require millions of qubits~\cite{gidney2021a,beverland2022}. In many physical computing architectures, it is challenging to imagine scaling a single device to this number of qubits, because of varied constraints including cryogenic cooling power, wiring density, or laser power. These challenges can be circumvented with a modular approach, using remote connections to link together a number of small units into a single quantum processor~\cite{meter2008, Monroe2014,nickerson2014,li2016}. Modularity may also simplify the construction, maintenance and calibration of large-scale systems.

The basic building block of a modular computer is a Bell pair between physical qubits in two modules, which can be used as a resource to teleport quantum states or gates between the modules~\cite{gottesman1999a,Olmschenk2009,pfaff2014,Chou2018}. However, it is an outstanding challenge to develop an interface between modules with sufficient bandwidth and fidelity to enable scalable, fault-tolerant computation. In superconducting qubits, proof-of-concept demonstrations have shown cryogenic microwave links between remote qubits~\cite{storz2023}, and entanglement between microwave and optical photons~\cite{sahu2023,meesala2023}. Remote entanglement of atomic qubits such as neutral atoms, trapped ions and solid-state defects using photons has been implemented with several approaches, using free-space and cavity-enhanced light-matter interfaces~\cite{moehring2007,Olmschenk2009,ritter2012,hofmann2012,bernien2013,Hucul2014,delteil2016,stephenson2020,vanleent2022,krutyanskiy2023,knaut2023}. However, the highest reported remote entanglement rate between neutral atom or trapped ion qubits is only 182 s$^{-1}$, with a fidelity of 0.94~\cite{stephenson2020}.

In this work, we propose an approach to realize fast, high-fidelity remote qubit entanglement between neutral atom quantum processors~\cite{bluvstein2023,graham2022a}. We consider $^{171}$Yb atoms as qubits, which have been used to demonstrate high-fidelity entangling gates~\cite{ma2023a}, non-destructive~\cite{huie2023a} and mid-circuit~\cite{lis2023a,norcia2023a} readout, and have a pathway to hardware-efficient fault-tolerant error correction~\cite{wu2022,sahay2023a}. By using an optical tweezer array to place $N>100$ atoms inside a single optical cavity (a twisted ring resonator), and controlling their interaction with the cavity using only local light shifts, we predict a remote Bell pair generation rate of $1.0\times 10^5$ s$^{-1}$ with fidelity of approximately $0.999$ for physically reasonable parameters. This rate is orders of magnitude higher than previously demonstrated~\cite{Hucul2014,hofmann2012,stephenson2020,krutyanskiy2023} or proposed~\cite{huie2021a,young2022a} remote entanglement approaches for atomic qubits. The entanglement is distributed between modules using 1389 nm photons in a single-mode optical fiber, allowing links over several kilometers without degradation. We conclude with a brief discussion of the implication of these results for large-scale neutral atom quantum processors.

\begin{figure*}
    \centering
    \includegraphics[width=\textwidth]{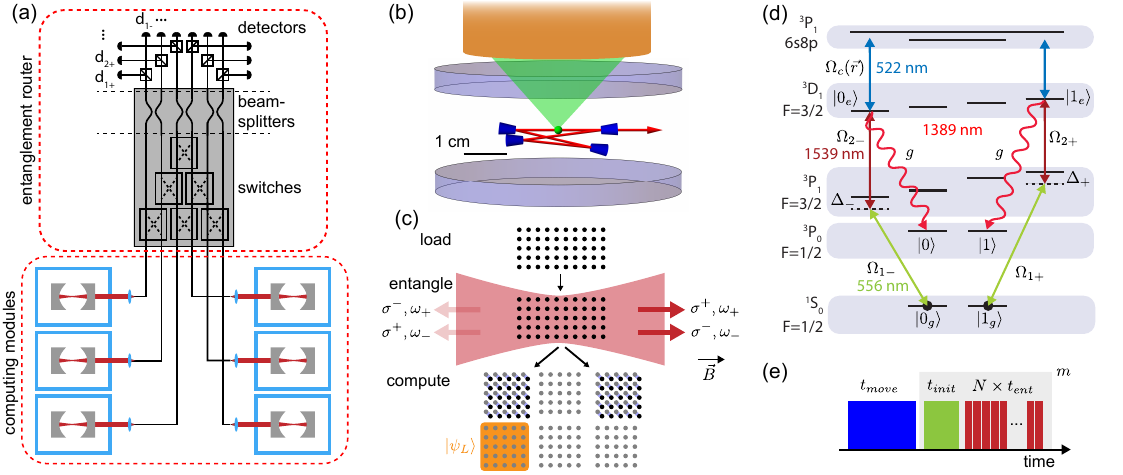}
    \caption{(a) Schematic of a modular neutral atom quantum computer with multiple computation modules and a central entanglement router. (b) Each module houses a centimeter-scale twisted ring cavity, shown here with vacuum windows and a high-NA microscope objective with standard dimensions for scale~\cite{Saskin2019}. (c) The entanglement process consists of loading and transporting arrays of atoms into the cavity, entangling them sequentially with a remote module, and moving them into a computation zone. There, they can be used to connect remote logical qubits $\ket{\psi_L}$ using teleported gate operations.
    (d) Energy diagram of relevant atomic levels in $^{171}$Yb. 
    The qubit state used for computation is in the $^3P_0$ manifold. 
    The atoms are excited from the $^1S_0$ ground states to the $^3D_1$ states with a two-photon excitation using $^3P_1$ as an intermediate state. The Rabi frequency of the excitation lasers are $\Omega_{1\pm}$ and $\Omega_{2\pm}$, with detuning to intermediate state of $\Delta_\pm$. The $^3D_1$ to $^3P_0$ transitions are coupled to the cavity with atom-cavity coupling strength of $g$.
    Local light shifts are is applied to selected atoms, using $\Omega_c(\vec r)$ coupling $^3D_1$ to a higher excited state.
    (e) Temporal sequence of remote entanglement generation: by amortizing the cost of moving and initializing atoms over many entanglement attempts, the overall attempt rate is very close to $1/t_{ent}$.}
    \label{fig:fig1}
\end{figure*}

\section{Overview}

An overview of the proposed approach is shown in Fig.~\ref{fig:fig1}. We envision an array of computation modules each housed in a separate vacuum system containing an optical cavity used to generate spin-photon entanglement. Photons are emitted on the 1389 nm transition from $^3D_1$ to $^3P_0$ ($\Gamma = 2 \pi \times 418$~kHz)~\cite{covey2019a}, with a polarization entangled with the qubit states in $^3P_0$. After the photons leave the optical cavity, they are coupled into optical fibers, and sent to a central router and detector array~\cite{Monroe2014}. The router interferes photons coming from selected pairs of modules on an array of beamsplitters.
The coincident detection of two photons on the same beamsplitter heralds the generation of a Bell state, with a success probability of 50\%.

The optical cavity within each module is a non-planar (twisted) ring cavity (Fig.~\ref{fig:fig1}b)~\cite{Schine2016nature}, which allows small beam waists with robust alignment~\cite{jia2016,cox2018,chen2022b}, spatially uniform atom-cavity coupling (\emph{i.e.}, without a standing wave), and non-degenerate modes of opposite circular polarization~\cite{jia2018} that are used to couple to two Zeeman-split transitions simultaneously to generate spin-polarization entanglement. This doubles the entanglement generation rate compared to time-bin entanglement when coupling to a single transition.

The operation of the modular interface with a tweezer array is shown in Fig.~\ref{fig:fig1}(c-e). 
An array of $N$ atoms is initialized in a loading zone, then transported into the cavity. Once inside, the entire array is initialized in a superposition state within the $^1S_0$ ground state $\ket{\psi_0}^{\otimes N} = [(\ket{0_g} + \ket{1_g})/\sqrt{2}]^{\otimes N}$. The first atom in the array is excited to a superposition state in the $6s5d\,^3D_1$ manifold, $\ket{\psi_e} = (\ket{0_e} + \ket{1_e})/\sqrt{2}$, which decays to the qubit state in the $^3P_0$ manifold, by emitting a 1389 nm photon into the cavity with $\sigma^-$ or $\sigma^+$ polarization. This results in the spin-photon entangled state $\ket{\psi_{sp}} = \left(\ket{0, \sigma^-} + \ket{1,\sigma^+}\right)/\sqrt{2}$. By performing the excitation synchronously in two modules, the emitted photons will arrive simultaneously at the detectors, and an entangled state of two qubits $\ket{\psi^\pm} = (\ket{01} \pm \ket{10})/\sqrt{2}$ is heralded when two photons of opposite polarization are detected~\cite{moehring2007a}.

The process is repeated sequentially for each atom in the array, with a delay of $t_{ent} \approx 1\,\mu$s to ensure that the photon wavepackets do not overlap. After exciting all of the atoms once, the procedure can be repeated $m \approx 5-10$ times to boost the entanglement fraction, by re-initializing the atoms that were not successfully entangled in previous rounds (Fig.~\ref{fig:fig1}e). After a sufficient number of Bell pairs are generated, the array is moved out of the cavity and replaced by a fresh array to continue generating more Bell pairs. By amortizing the relatively high temporal cost of moving ($t_{move} \approx 100\,\mu$s~\cite{bluvstein2022}) and initialization ($t_{init} \approx 6-8\,\mu$s) over many repetitions of the entanglement sequence, the average spin-photon entanglement attempt rate approaches the maximum allowed by the cavity, $1/t_{ent}$. Saturating this rate requires $N \gtrsim (t_{move}+t_{init})/t_{ent} \approx 100$. 
Locally addressed light shifts of the $^3D_1$ state prevent reabsorption of photons by atoms already in $^3P_0$.

The subset of qubits that were successfully entangled can be used to perform remote operations between logical qubits $\ket{\psi_L}$ in two modules, for example by using teleported gates to implement lattice surgery or transversal entangling gates.

\section{Entanglement rate}

\begin{figure}
    \centering
    \includegraphics[width=\columnwidth]{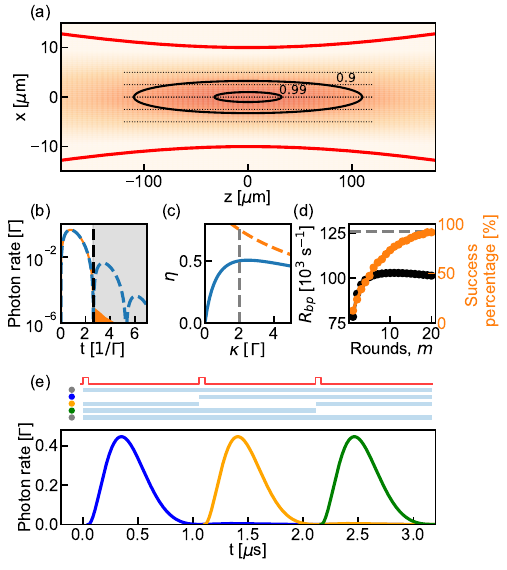}
    \caption{Entanglement rate. 
    (a) Schematic of an atom array inside the cavity. The red contour indicates the $1/e^2$ intensity of the cavity field, while the black lines show $g=0.99g_{max}$ and $g=0.9g_{max}$. The dots depict an atom array with a spacing of 2.5\,$\mu$m. $N=204$ sites fit inside the $0.9 g_{max}$ contour.
    (b) Output intensity from the cavity after exciting an atom to $^3D_1$ (blue dashed line). A light shift $\Delta$ is applied at $t \approx \pi/g = 2.65/\Gamma$ suppresses the late time decay (orange line, $\Delta/g = 100$), allowing the next atom to be excited immediately.
    (c) Probability of emission into the cavity, $\eta$, as a function of cavity decay rate, $\kappa$ (blue). The orange curve shows the maximum possible value, $\eta_0 = C/(C+1)$. The optimum $\kappa$ is indicated by the vertical line, where $\eta = 0.50$.
    (d) Bell pair generation rate $R_{bp}$ and the percentage of atoms entangled, as a function of the number of rounds, $m$.
    (e) Timing diagram for the entanglement sequence in a small array, showing the local light shifts (blue bars), global excitation pulses (red) and cavity output field. Note that at each point in time, a light shift is applied to all atoms \emph{except} for the one being entangled. 
    }
    \label{fig:fig2}
\end{figure}

An example cavity design is shown in Fig.~\ref{fig:fig1}b, consisting of four mirrors in a twisted ring. The large spacing between the mirrors and the atoms eliminates unwanted atom-surface interactions and provides ample optical access for optical tweezers, imaging and gate beams. The twisted geometry breaks the degeneracy of modes with opposite helicity~\cite{jia2018}, resulting in a controllable splitting between co-propagating modes with $\sigma^+$ and $\sigma^-$ polarization. The twist angle of $\approx 11^\circ$ is chosen to tune the forward-propagating modes into resonance with the transitions from $\ket{0}$ and $\ket{1}$ to $^3D_1$ simultaneously, in a bias field of 100 Gauss ($\omega_+ - \omega_- = 140$~MHz). The mirror radius of curvature and arm lengths are chosen to provide a circular mode with a $1/e^2$ waist $w_0 = 10\,\mu$m at the position of the atoms. This results in a peak atom-cavity coupling strength of $g_{max} = 2\pi \times 520$\,kHz for both transitions, and provides space for an array of $N=204$ atoms with $g > 0.9 g_{max}$ (Fig.~\ref{fig:fig2}a). The round-trip length is $L = 6.96$\,cm, corresponding to a free spectral range of 4.3\,GHz, and the cavity decay rate $\kappa$ is chosen by selecting the reflectivity of the outcoupler mirror (the other mirrors have $R=1$). Additional details about the cavity design can be found in Appendix~\ref{sec:cavityparameters}.

The procedure for generating the spin-photon entangled state $\ket{\psi_{sp}}$ is as follows. 
After preparing the array in $\ket{\psi_0}^{\otimes N}$, we use a two-photon excitation pulse with Rabi frequency $\tilde{\Omega} = \Omega_{1\pm}\Omega_{2\pm}/2\Delta_{\pm}$ to excite a single atom to $\ket{\psi_e}$.
The atom decays by emission into the cavity and free space modes, and the cavity output is shown in Fig.~\ref{fig:fig2}b.
To avoid residual population in the cavity causing an error on the next entanglement attempt, one would have to wait for a time $t>5 / \Gamma$.
However, the exponential tail can be suppressed by tuning the atom out of resonance with the cavity suddenly at $t=\pi/g$, when the cavity population vanishes, using a strong light shift $\Delta$ on the $^3D_1$ state with $\Delta/g > 100$. This traps the residual excitation in the atom, where it decays into free-space modes, and allows the next atom to be excited to $^3D_1$ immediately, doubling the emission rate to $t_{ent} = 2.65/\Gamma$, while decreasing the emission probability per attempt by only 1\%. The light shift is kept on for the remainder of the sequence, to prevent atoms in $^3P_0$ from absorbing cavity photons from subsequent entanglement attempts. After attempting to entangle all $N$ atoms, the atoms that were not entangled can be repumped back to $\ket{0_g}$, and the entire procedure repeated for another round. 

Fig.~\ref{fig:fig2}c shows the probability of emission into the cavity as a function of $\kappa$, reaching a maximum of $\eta=0.50$ when $\kappa/2\pi = 1.04$~MHz (corresponding to a cavity finesse of $\mathcal{F} = 2070$). This does not achieve the maximum possible efficiency $\eta_0 = C/(1+C)$, where $C = 4 g^2/(\kappa \Gamma) \approx 2.5$ is the cooperativity, because the cavity becomes spectrally narrower than the emitted photon when $\kappa$ is small~\cite{young2022a}. Adiabatic preparation of a shaped photon pulse can saturate $\eta_0$~\cite{vasilev2010} by increasing $t_{ent}$, but we have found that this does not increase the remote entanglement rate and adds significant additional experimental complexity.

Each entanglement attempt heralds a successful Bell pair with probability $P_{suc} = (1/2) \eta^2 = 0.125$. Using the sequence in Fig.~\ref{fig:fig1}e, the average Bell pair rate when using $m$ rounds is:
\begin{equation}
\label{eq:Re}
    R_{bp} = \frac{\sum_{i=1}^m N_i P_{suc}}{t_{move} + m\cdot t_{init} + \sum_{i=1}^m N_i \bar{t}_{ent}}
\end{equation}
where $N_i = N_{i-1}(1-P_{suc})$ is the number of entanglement attempts in round $i$ ($N_1 = N$), and $\bar{t}_{ent} = 1.09\,\mu$s is the average time per entanglement attempt across the $N=204$ site array. After one round ($m=1$) the rate is $7.8 \times 10^{4}$\,s\,$^{-1}$, increasing to $1.0\times 10^5$\,s\,$^{-1}$ for $m=5-20$ rounds. This is 82\% of the maximum rate allowed by the cavity, $1/t_{ent} = 1.25 \times 10^{5}$\,s\,$^{-1}$, indicating the effectiveness multiplexing the cavity across $N$ atoms.

The same light shift control can also be used to address the excitation and repumping operations, as described in Appendix~\ref{sec:experimentsequence}. A timing diagram of part of one round is shown in Fig.~\ref{fig:fig2}e.

\section{Entanglement Fidelity}

\begin{figure}
    \centering
    \includegraphics[width=\columnwidth]{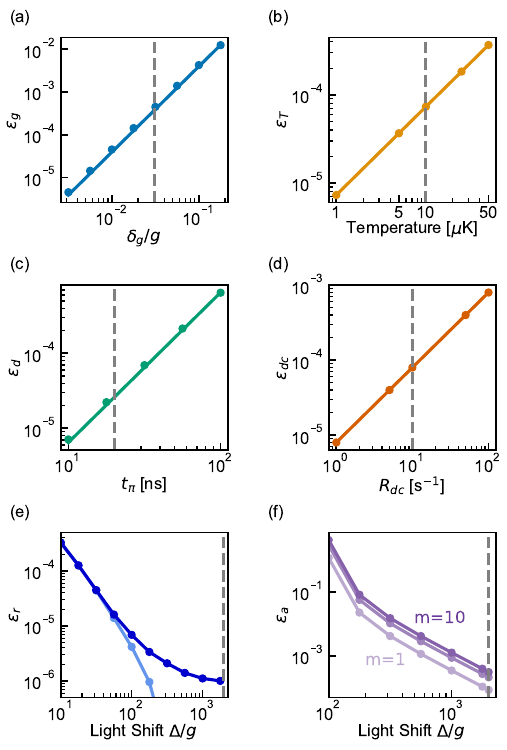}
    \caption{(a) Error resulting from unequal atom cavity coupling strength, $\epsilon_g$. The dashed line shows a representative value, $\delta_g/g = 0.031$, derived from the r.m.s. variation across the array shown in Fig.~\ref{fig:fig2}a.
    (b) Doppler shift error, $\epsilon_T$, from finite atomic temperature. The dashed line shows a typical temperature for Yb atoms after Doppler cooling.
    (c) Error from re-excitation of the atom following a decay during the excitation pulse, $\epsilon_d$. 
    (d) Error from detector dark counts, $\epsilon_{dc}$.
    (e) Error probability from residual cavity photon population after each entanglement attempt, $\epsilon_{r}$, as a function of the light shift detuning $\Delta/g$. The light line shows the case that $g$ is known exactly, while the dark line includes an uncertainty in $g$ of $\delta_g/g = 3 \times 10^{-3}$, corresponding to thermal atomic motion at $T=10\,\mu$K.
    (f) The error probability $\epsilon_{a}$ from scattering or Stark shifts from cavity photons in subsequent rounds, as a function of the light shift detuning $\Delta/g$. Purple lines show the error after $m=1,5,10$ rounds (from light to dark).
    }
    \label{fig:fig3}
\end{figure}

Next, we consider errors affecting the fidelity of the remote Bell pairs. Previous experimental studies of heralded entanglement based on coincidence detection have identified errors arising from qubit decoherence, the fidelity of single-qubit rotations, polarization mixing and imperfect mode overlap at the detectors~\cite{Hucul2014,stephenson2020}.

The latter effects are strongly suppressed by the cavity. The mode structure of the cavity ensures that the photons are emitted into a single, well-defined mode with orthogonal polarizations, which can be preserved through the entanglement router by using fiber optic or integrated photonic beamsplitters with extremely low loss at this wavelength~\cite{timurdogan2019}.

Furthermore, the excellent coherence and high-fidelity single-qubit operations on the nuclear spin qubit in $^{171}$Yb largely mitigates the first two effects. Coherence times without dynamical decoupling (\emph{i.e.}, $T_2^*$) of several seconds have been demonstrated for the pure nuclear spin qubit in $^1S_0$ and $^3P_0$, because of the low sensitivity to magnetic field noise and absence of differential light shifts~\cite{ma2022,jenkins2022,ma2023a,lis2023a}. The $^3P_0$ state lifetime is 3\,s in typical optical tweezers~\cite{ma2023a}, resulting in a decay probability to $^1S_0$ of $7 \times 10^{-7}$ per $t_{ent}$ or $2 \times 10^{-4}$ over $m=5$ sequence repetitions with $N=204$ atoms. We also note that $^3P_0$ decays can be converted into erasure errors, which can be efficiently corrected~\cite{wu2022,ma2023a}. Unlike $^3P_0$ and $^3S_0$, the $^3D_1$ state is sensitive to magnetic fields. However, it is only populated for $\approx 1\,\mu$s during the entanglement generation, compatible with phase accumulation errors less than $10^{-6}$ using conventional field stabilization at the part-per-million level~\cite{merkel2019}. Single qubit gates for both $^1S_0$ and $^3P_0$ qubits have been demonstrated with fidelities beyond 0.999~\cite{ma2022,jenkins2022,ma2023a,lis2023a}.

Now, we turn to some sources of error that are intrinsic to our implementation. The magnitude of these error sources is estimated from numerical simulations (Appendix~\ref{sec:sims}) unless otherwise stated. The first is a slight distinguishability of the photons from two atoms resulting from variations in the atom-cavity coupling strength. Given two atoms in two cavities with a fractional difference in coupling strength $\delta_g/g$, the resulting distinguishability causes an error $\epsilon_g = 0.394\times (\delta_g/g)^2$, such that $\delta_g/g < 0.05$ is required to reach $\epsilon_g < 10^{-3}$ (Fig.~\ref{fig:fig3}a). Maintaining an r.m.s. variation in $g$ below this level for all possible qubit pairs requires matching the cavity waists within 5\%, and placing the atoms within $0.4 w_0 \approx 4\,\mu$m of the cavity axis (compatible with the layout in Fig.~\ref{fig:fig2}a). Static inhomogeneity in $g$ can also be mitigated by misaligning the atoms in one cavity, or choosing matched pairs of atoms to entangle. 
However, unknown variation in $g$ can arise from thermal motion of the atoms or alignment drifts between the tweezer array and the cavity. Thermal motion at $10\,\mu$K corresponds to $\delta_g/g \approx 3 \times 10^{-3}$, while maintaining $\delta_g/g < 0.05$ requires alignment stability to within $(\Delta x, \Delta y, \Delta z) < (0.87, 2.2, 3.8)\,\mu$m, compatible with demonstrated tweezer alignment stability to nanophotonic structures~\cite{tiecke2014} and standing wave lattices~\cite{schine2022}.

Distinguishability errors can also arise from Doppler shifts, given the running-wave mode in the cavity. The error probability $\epsilon_T$ is proportional to $(k_c v_{rms} / g)^2$, where $k_c = 2\pi/\lambda$ is the cavity wavevector, and $v_{rms} = \sqrt{k_B T/m}$ is the mean atomic velocity along the cavity axis (with $k_B$, $T$ and $m$ denoting Boltzmann's constant, the atomic temperature and the mass of the atom, respectively). From numerical simulations, we find $\epsilon_T = 7.3\times 10^{-6} \mu \textrm{K}^{-1} \times T$ (Fig.~\ref{fig:fig3}b). A typical temperature for Yb atoms after Doppler cooling is 5-10 $\mu$K~\cite{Saskin2019,jenkins2022,ma2023a}, corresponding to $k_c v_{rms} = 2 \pi \times 16$ kHz, yielding $\epsilon_T < 10^{-4}$.

Next, we consider decay back to $^1S_0$ during the excitation pulse, which collapses the initial spin superposition but can still result in photon emission into the cavity if the atom is re-excited. The $^3D_1$ state decays to $^3P_1$ with a branching ratio of 0.34, and $^3P_1$ decays to $^1S_0$ at a rate $\Gamma_3 = 2\pi \times 182$~kHz. Therefore, the probability to decay to $^1S_0$ during an excitation pulse of duration $t_\pi \ll \Gamma, \Gamma_3$ is $p_3 \approx 0.34 t_\pi^2 \Gamma_3 \Gamma$. The probability of the atom being re-excited and causing an error is $\epsilon_d \approx p_3/8$. From numerical simulations, we find $\epsilon_d = 6.8\times10^{-8}$ns$^{-2}$$\times t_{\pi}^2$, which yields $\epsilon_d = 2.8\times 10^{-5}$ for $t_\pi = 20$~ns. In Appendix~\ref{sec:experimentsequence}, we discuss errors from unwanted excitation to $^3D_1, m_F =\pm1/2$ levels and off-resonant scattering from $^3P_1$, which can both be suppressed below this level.

Then, we consider false heralding events from detector dark counts (Fig.~\ref{fig:fig3}d). The probability of a false herald is $p_f = 2\eta R_{dc}t_{ent}$, where $R_{dc}$ is the dark count rate on a single detector. Therefore, the probability of an error caused by dark count is $\epsilon_d = p_f/P_{suc} = 4 t_{ent}R_{dc}/\eta \approx 8\times 10^{-6}$~s~$\times R_{dc}$. Commercially available SNSPDs have a dark count rate $R_{dc} = 10$~s\,$^{-1}$, corresponding to an error probability of $\epsilon_{dc} = 8\times 10^{-5}$. Suppression to millihertz rates has been demonstrated using cold filtering~\cite{shibata2015} or integrating the detector to a single-mode waveguide~\cite{schuck2013}.

Finally, we consider errors from the presence of multiple atoms inside the cavity. These errors take two forms: residual photons in the cavity at the end of one entanglement attempt leaking into the next window, and atoms already in $^3P_0$ absorbing a photon if they are not sufficiently light-shifted away from the cavity resonance (Fig.~\ref{fig:fig3}f). 
If the atom-cavity coupling is known exactly, the residual photon errors are suppressed by the light shift as $\epsilon_r \propto (\Delta/g)^{-2}$ , provided the light shift is turned on instantaneously at the time when the cavity population is zero (Fig.~\ref{fig:fig3}e). However, unknown variation in $g$ makes this precise timing impossible. Using the uncertainty $\delta_g/g = 3\times 10^{-3}$ from thermal atomic motion at $10\,\mu$K, we find a floor of $\epsilon_r < 10^{-5}$ with a light shift of $\Delta/g > 100$.

Adding the above sources of error, we arrive at an average Bell pair fidelity of approximately 0.999, dominated by $\epsilon_a$ and $\epsilon_g$.

Atoms already in $^3P_0$ after a successful entanglement attempt can experience scattering or Stark shifts from photons in the cavity during subsequent entanglement attempts. Both errors are suppressed with the light shift as $(\Delta/g)^{-2}$, and we find from numerical simulations that the error probability after one additional entanglement attempt is $\epsilon_a^{(1)} = 3.5\times (\Delta/g)^{-2}$. However, because the error is incurred on the atom that was already entangled, the average generated Bell pair will experience $\bar{N}_r = \mathcal{O}(m N)$ subsequent entanglement attempts before being transported out of the cavity, amplifying its effect. For $N=204$ and $m=5$, we find numerically that $\bar{N}_r \approx 250$. Achieving an average error $\epsilon_a = \bar{N}_r \epsilon_a^{(1)} < 10^{-3}$ requires $\Delta/g > 2 \times 10^3$. 

We note that only a modest amount of laser power is required for the light-shifting beam when operating close to resonance on the $^3D_1$ to $6s8p\, ^3P_1$ transition, which is possible because the $^3D_1$ state is not populated on the shifted sites, and therefore insensitive to scattering errors as demonstrated in Ref.~\cite{burgers2022}. With a detuning of 1 GHz, a laser power of approximately $20\mu$W per atom is sufficient to generate a light shift of $\Delta/g>2\times 10^3$, with negligible mechanical forces on the atoms as the wavelength is far from resonance with any transitions from $^1S_0$ or $^3P_0$.
Unwanted light shifts on the atom being entangled can arise from crosstalk of the local addressing beam, and lead to distinguishablity errors similar to Doppler shifts. A recently demonstrated large-scale modulator achieves -44 dB nearest-neighbor crosstalk~\cite{zhang2023a}, which would give a detuning error $2.6$ times the r.m.s. Doppler shift, or an entanglement error of $5\times 10^{-4}$.

\subsection{Information about errors from photon timing}

\begin{figure}
    \centering
    \includegraphics[width=\columnwidth]{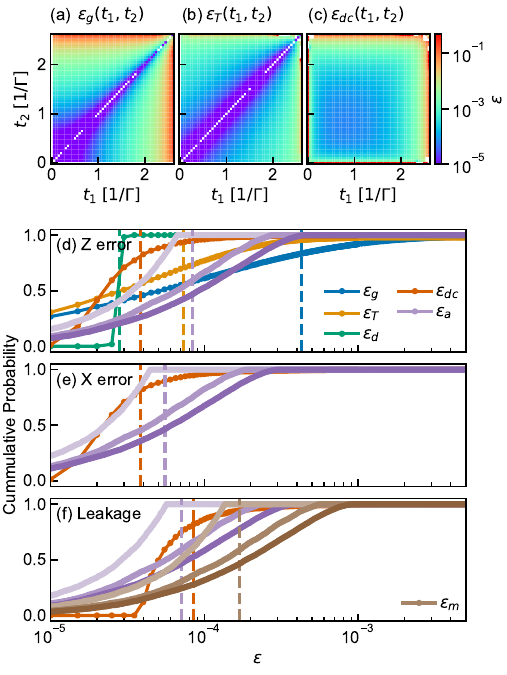}
    \caption{(a) Probability of an error from unequal $g$, as a function of the photon detection time $(t_1,t_2)$. (b) Probability of an error from Doppler shifts as a function of detection time. (c) Probability of a dark count error as a function of photon detection time.
    (d-f) Cumulative probability distribution functions for the $Z$, $X$ and leakage error rates for all error sources considered. The dashed lines show the mean probability of each error. For $\epsilon_a$ and $\epsilon_m$, the (light, medium, dark) curves show the error probability using $m=(1,5,10)$ rounds of entanglement generation, and the dashed line shows the mean error for $m=5$. For all error types, the strength of the noise corresponds to the dashed vertical lines in Fig.~\ref{fig:fig3}, and $\epsilon_m$ is computed based on a 3 s lifetime for $^3P_0$~\cite{ma2023a}.
    }
    \label{fig:fig4}
\end{figure}

In the context of quantum error correction, the details of the error model can significantly affect the overhead required to reach a given logical error rate. For example, a bias towards a single type of Pauli error~\cite{Aliferis2008,BonillaAtaides2021} or information about the location of errors in the form of erasures~\cite{grassl1997,stace2009,wu2022,sahay2023a} or soft information~\cite{pattison2021} have been shown to reduce logical error rates by several orders of magnitude. While many quantum operations can only be characterized by their average fidelity, here we show that the timing of the photon detection events provides shot-to-shot information about the probability of different error types of each individual Bell pair. 

Specifically, we ask the question: given that two photons are detected at times $(t_1,t_2)$, what is the probability that the resulting qubit state has a Pauli or leakage error? Since the qubit manifold in $^{171}$Yb has only two sublevels, leakage refers to the case that the atom is not in $^3P_0$.

First, we consider the errors caused by photon distinguishability from a variation in the atom-cavity coupling strength. This results in unequal wavepacket shapes for the two photons, which encodes which-path information into the detection time difference. These errors are purely Pauli $Z$ errors, as the correlation between the photon polarization and the qubit state in the computational basis is not affected. In Fig.~\ref{fig:fig4}a, we plot the error probability as a function of the photon detection times, $\epsilon_g(t_1,t_2)$. Using the probability to detect photons at those times, $P(t_1,t_2)$, we can convert $\epsilon_g(t_1,t_2)$ into a probability distribution for the error rate of the generated Bell pairs, $p(\epsilon_g)$, whose cumulative distribution function is shown in Fig.~\ref{fig:fig4}d. The function $p(\epsilon_g)$ gives the probability that a particular Bell pair has a fidelity $1-\epsilon_g$. While the mean error rate is $4\times 10^{-4}$, over half of the Bell pairs have an error rate less than $6\times 10^{-5}$ . This  approximates an erasure error and allows for better decoding and improved logical qubit performance.

A similar analysis applies to errors from Doppler shifts. The probability of error as a function of the photon detection times, $\epsilon_T(t_1,t_2)$ is shown in Fig.~\ref{fig:fig4}b. As Doppler shifts also only affect the distinguishability of the photons, this only results in $Z$ errors, and is suppressed when $\left|\Delta \omega (t_1-t_2)\right| \ll 1$, where $\Delta \omega$ is the frequency difference of the two atoms~\cite{zhao2014}. The cumulative distribution for $p(\epsilon_T)$ is shown in Fig.~\ref{fig:fig4}d.

Errors from decay during the excitation pulse, $\epsilon_d$, can be understood as a Z-basis measurement of the atom prior to the spin-photon entanglement, and are also purely $Z$ errors. However, the photon detection timing does not reveal anything about the probability of this error, so it is the same for all Bell pairs and the the cumulative  distribution is a step function (Fig.~\ref{fig:fig4}d).

Dark counts result in false heralding signals that are uncorrelated with the atomic state, and can therefore cause all Pauli errors and leakage. The error probability $\epsilon_{dc}(t_1,t_2)$ depends on the detection time, and is greatest at the edge of the detection window when the probability of detecting a real photon is low (Fig.~\ref{fig:fig4}c). The corresponding cumulative distribution functions for $Z$, $X$ and leakage errors is shown in Fig.~\ref{fig:fig4}d-f.

Finally, we consider errors from the absorption of photons by atoms already in Bell pairs, $\epsilon_a$. 
Because this error happens after the Bell pair is created, it does not depend on the photon detection times. However, Bell pairs created in early rounds have more chances to experience an error than Bell pairs from later rounds. Using the known probability distribution for the number of entanglement attempts $N_r$ that follow a successful Bell pair creation, we can generate the probability distribution $p(\epsilon_a)$ for $Z$, $X$ and leakage errors (Fig.~\ref{fig:fig4}d-f). Because the atom is detuned far from the cavity, the probabilities to excite to all $m_F$ levels of $^3D_1$ are comparable, and therefore the probability of $X$, $Z$ and leakage errors are also roughly comparable. Errors from the finite $^3P_0$ lifetime, $\epsilon_m$, are purely leakage, and follow a similar probability distribution (Fig.~\ref{fig:fig4}f).

In summary, we have shown that the fidelity of individual Bell pairs is strongly influenced by information that is available to the experimenter: the photon detection times, and the time step in the sequence when the Bell pair was generated. Passing this information to the decoder when performing logical operations may significantly reduce the logical error rate~\cite{wu2022,pattison2021}. We leave a detailed analysis to future work.

\section{Discussion}

We conclude by discussing several aspects of these results. First, we note that the proposed approach is fully compatible with existing neutral atom quantum processors. The cavity allows for a large separation between the atoms and the mirror surfaces, comparable to standard glass cell vacuum chambers~\cite{Saskin2019}. It also provides large optical access, compatible with high-numerical-aperture objective lenses for projecting tweezer arrays and local addressing beams, and for fluorescence imaging. Moreover, the steps involved in generating entanglement can proceed in parallel with computations using metastable qubits in a nearby zone, without the need to separately control magnetic fields. The local addressing requirement is minimal, consisting of only one switchable light shift on each site, and is compatible with recently demonstrated scalable modulators~\cite{menssen2023,zhang2023a}.

Second, we consider how physical Bell pairs can be used to implement fault-tolerant operations between remote logical qubits. Previous studies have considered modular quantum computing in the regime where the modules are small, and a single logical qubit spans multiple modules~\cite{jiang2007,moehring2007a,Monroe2014,li2016}. While the existence of a high error threshold of over 10\% has been demonstrated for the inter-module links~\cite{nickerson2014,li2016}, remote Bell pairs are consumed at a high rate just to sustain the logical information against idle errors.

In the case of neutral atom quantum computing, we envision modules with $\mathcal{O}(10^4)$ qubits per module, based on demonstrated arrays of hundreds of qubits~\cite{ebadi2021,scholl2021} and scaling of the underlying optical components beyond $10^4$ sites~\cite{zhang2023a}. With foreseeable error rates below $10^{-3}$ for all physical operations~\cite{ma2023a,scholl2023a,evered2023,lis2023a}, achieving logical error rates of $10^{-12}$ is possible with an overhead of $10^3$ physical qubits per logical qubit using a standard surface code~\cite{gidney2021b}. The overhead may be reduced by at least another order of magnitude using qubits engineered for erasure-biased errors~\cite{wu2022,sahay2023a} or efficient block codes~\cite{xu2023}. Therefore, $N_L \gtrsim 100$ logical qubits per module is realistic. In this regime, the remote Bell pairs are not used to correct idle errors and errors from local operations within each module, but only for performing logical operations between modules.

For a distance $d$ surface code, a gate operation between remote logical qubits can be implemented via lattice surgery~\cite{horsman2012}, consuming $d^2$ physical Bell pairs to teleport CNOT gates along the code boundary~\cite{ramette2023}. Alternatively, the same number of Bell pairs can be used to implement a logical transversal CNOT across two modules, by teleporting CNOT gates between corresponding physical qubits in the code. The second approach can also be applied to $[[n,k,d]]$ quantum LDPC codes~\cite{breuckmann2021} with  a transversal CNOT$^{\otimes k}$ gate to generate $k$ entangled logical qubits between two modules by consuming $n$ physical Bell pairs, allowing a higher rate of logical Bell pair generation than the surface code. The implementation of such codes for neutral atom qubits has been discussed in Ref.~\cite{xu2023}. As the fidelity of remote physical Bell pairs is similar to those that could be created within a module, intermediate purification steps before using the Bell pairs in logical operations are not required.

Therefore, the physical interface proposed here will enable $\gtrsim 10^3$ remote logical gates per second between each pair of modules. It is an interesting question for future work to consider how to compile high-level algorithms into such a modular computer~\cite{meter2008,Monroe2014,gidney2021a}.

\section{Conclusion}

We have presented a blueprint for a modular architecture for a neutral atom quantum processor based on Yb atoms. It is capable of generating remote entanglement at a rate of $1.0 \times 10^{5}$ Bell pairs per second, with a fidelity  compatible with fault-tolerant computing. The modular interface can be implemented with existing experimental hardware, and  operated alongside an atom array performing local computations.

\section{Acknowledgements}

We gratefully acknowledge Jon Simon for sharing details about computing the normal modes of twisted resonators. We also acknowledge Nathan Schine, Mehmet Uysal, Shruti Puri for helpful discussions, and Adam Kaufman for comments on the manuscript. This work was supported by the Gordon and Betty Moore Foundation, grant DOI 10.37807/gbmf12253.

\appendix

\section{Experimental sequence}
\label{sec:experimentsequence}

\begin{figure}
    \centering
    \includegraphics[width=\columnwidth]{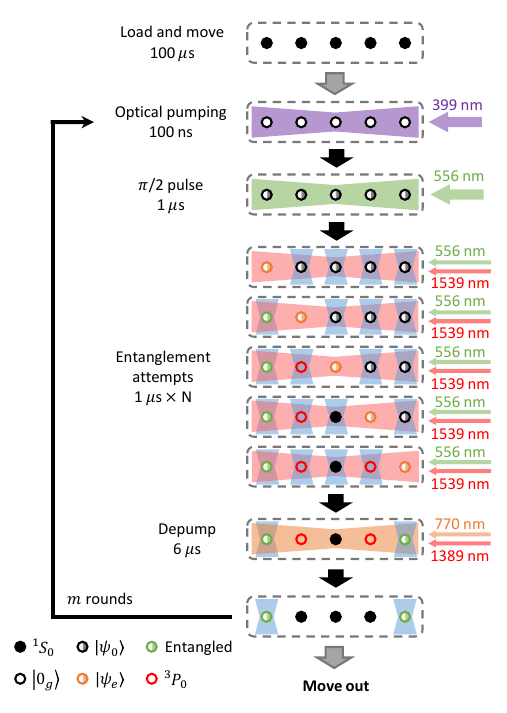}
    \caption{Schematic diagram of the experimental sequence on a five atom array. }
    \label{fig:appen_timing}
\end{figure}

A detailed description of the experimental sequence is shown in Fig.~\ref{fig:appen_timing}. Here, we discuss each of the steps in additional detail.

After moving atoms into the cavity, the entanglement sequence begins with initializing the array to state $|\psi_0\rangle$. This has two steps: optical pumping to $|0_g\rangle$ on the $^1S_0$ to $^1P_1$ transition at 399 nm (requiring less than 100~ns), and a Raman $\pi/2$ rotation ($~170$~ns~\cite{jenkins2022}). We note that these steps do not affect qubits in $^3P_0$, so can be used in subsequent rounds without affecting Bell pairs that have already been created.

Then, the atoms are sequentially excited to $^3D_1$, using the laser configuration in Fig.~\ref{fig:fig1}d. It consists of three laser beams: A $\sigma_+$-polarized laser driving $|1_g\rangle$ to the $^3P_1, m_F=+3/2$ state with detuning $\Delta_+$, a $\sigma_-$-polarized laser driving $|0_g\rangle$ to the $^3P_1, m_F=-3/2$ state with detuning $\Delta_-$, and a $\pi$-polarized laser driving the $^3P_1$ states to $|0_e\rangle,\,|1_e\rangle$. Simultaneous two-photon resonance is achieved for both transitions by controlling the frequency of the $\sigma^\pm$ lasers separately. The two photon Rabi frequency is given by $\tilde{\Omega}_\pm = \frac{\Omega_{1\pm}\Omega_{2\pm}}{2\Delta_{\pm}}$. The same laser configuration will also drive off-resonant Rabi oscillations to the $^3D_1, m_F = \pm 1/2$ states, with Rabi frequency:

\begin{equation}
    \tilde\Omega_\pm^\prime = \sqrt{(g_D\mu_B B)^2 + \left (\frac{1}{3\sqrt{3}}\frac{\Omega_{1\pm}\Omega_{2\pm}}{2(\Delta_\pm \pm g_P \mu_B B)}\right )^2},
\end{equation}
where $g_P = 1$ is the $^3P_1,F=3/2$ state Land\'e $g$ factor, $g_D = 1/3$ is the $^3D_1,F=3/2$ state Land\'e $g$ factor, and $\mu_B$ is the Bohr magneton. To minimize off-resonance excitation, the excitation pulse time $t_{\pi}$ should be an integer multiple of the off-resonance Rabi cycle, $t_{\pi} =\pi/\tilde \Omega_\pm = l\cdot 2\pi/{\tilde\Omega_\pm^\prime}$, where $l$ is an integer. For $\Delta_{\pm} = 2\pi \times 1$~GHz, and $B=100$~G, we
numerically found that the minimum pulse time approximately satisfying this condition is $t_{\pi} \approx 20$ ns. It cannot be exactly satisfied because $\tilde{\Omega}'_+ \neq \tilde{\Omega}'_-$, but the resulting excitation probability to $m_F=\pm1/2$ is only $3\times 10^{-6}$. This $t_\pi$ requires $\Omega_{1\pm} = \Omega_{2\pm}\approx2\pi\times 225$~MHz. The large detuning suppresses scattering from $^3P_1$ during the pulse to $\Gamma_3/\Delta_\pm \approx 10^{-4}$.

Instead of locally addressing the excitation lasers, which atom is excited can be controlled using a local light shift on $^3D_1$ during the excitation pulse to shift all but one atom out of resonance. This introduces several new sources of error, which we discuss only briefly because they are not intrinsic (\emph{i.e.}, they can be eliminated by locally addressing at least one of the excitation beams). As discussed in the main text, a value $\Delta/g = 2000$ (corresponding to $\Delta \approx 2\pi \times 1$~GHz) is needed to avoid cross-talk between atoms in the cavity. This is sufficient to suppress unwanted excitation: the probability of a scattering error from $^3D_1$ during the excitation pulse on a shifted site is approximately $t_\pi \Gamma (\tilde{\Omega}/\Delta)^2 = 3 \times 10^{-5}$. However, both this error and the scattering error from the $^3P_1$ intermediate state will accumulate, and are enhanced by $N$ for the final atom in each round. This can be suppressed by periodically re-initializing the atoms remaining in $^1S_0$ in the middle of each round, which adds little temporal overhead but will eventually heat the atoms from scattering optical pumping photons.

Even on light-shifted sites, the excitation laser can cause vacuum-stimulated Raman transitions resulting in the emission of a photon into the cavity modes. This coherent process occurs at a rate $g \Omega/\Delta \ll \Omega$, resulting in an emission probability of $(g/\Delta)^2 \approx 2 \times 10^{-7}$ per atom during the excitation pulse. This will have the same effect as a dark count or residual photons in the cavity from the previous round. Even when this probability is enhanced by the $N \approx 200$ $^1S_0$ atoms in the array, it remains well below the other sources of error discussed in the main text.

At the end of the each round, the atoms that are not in Bell pairs need to be depumped from $^3P_0$ before a new round can start. This can be done conveniently using the $^3D_1$ state, which decays to the ground state $^1S_0$ via $^3P_1$. We estimate that 6~$\mu$s is required to reach less than $10^{-3}$ remaining population in $^3P_0$. About $3\%$ of the population will decay to $^3P_2$, which is repumped using the $^3P_2$ to $^3S_1$ transition at 770 nm~\cite{ma2023a}. The $^3D_1$ light shift can also be used to control which atoms are depumped. Pumping of unwanted atoms is suppressed by $(\Delta/\Gamma)^2 \approx 10^{-6}$, as this error is incurred at most $m = 5-10$ times, it remains at a tolerable level.

At the end of a round, the entangled atoms are transported out of the cavity and into a computation zone with other logical qubits. At the same time, a new array is loaded into the cavity, so the entanglement sequence begins again after $t_{move}$.

To estimate the magnitude of the optical power needed to generate a light shift of $\Delta = 2000 g$, we use the experimentally measured lifetime of the $6s8p$~$^3P_1$ state of 140(20)~ns~\cite{bowers1996}, and an estimated branching ratio of 0.3 to the $^3D_1$ state. Using a detuning of $1$~GHz~\cite{burgers2022}, we arrive at an estimate of 20~$\mu$W/site, assuming a $1/e^2$ beam radius $w_0 = \lambda$. Several other excited states could be used for light shifts, including the $6s7p$ states (730 nm transition from $^3D_1$) or $6s5f\,^3F_2$ (528 nm). The latter state has a stronger matrix element to $^3D_1$ (26~ns lifetime~\cite{bowers1996}, estimated branching ratio of 90\% to $^3D_1$), which could reduce the needed power to approximately 1.5~$\mu$W per site.

\section{Cavity parameters}
\label{sec:cavityparameters}
The twisted, running-wave cavity in this work accomplishes two objectives. First, the running-wave nature reduces fast gradients in the atom-cavity coupling strength, enabling higher-fidelity spin-photon entanglement. Second, twisting the cavity gives rise to circularly polarized eigenmodes and lifts the degeneracy between the co-propagating modes with $\sigma^+$ and $\sigma^-$ polarization, by an amount that can be controlled by the twist angle. This allows maximum strength coupling to both $\sigma^+$ and $\sigma^-$ transitions simultaneously, allowing spin-polarization entanglement to be generated in the time it takes to emit one photon.

The design of the cavity is based on several principles. First, we constrain the atom-mirror distance to be larger than 1 cm to avoid deleterious effects on the Rydberg states of atoms near the mirror surfaces. This distance is comparable to the atom-windows separation in many current experiments~\cite{Saskin2019}. We also want a small mode waist, to realize a large atom-cavity coupling strength, and a twist angle that enables a splitting of 100-150 MHz between $\sigma^+$ and $\sigma^-$ modes in the same propagation direction, which is compatible with bias magnetic fields in the range of 60-100 G.

One example of a cavity satisfying these conditions is shown in Fig.~\ref{fig:fig1}b. It is made of four mirrors: two convex mirrors with radius of curvature $R=1.27$ cm, and two concave mirrors with $R=-1.27$ cm. The geometry is derived from a planar ring cavity with equal short and long arm lengths, and an opening angle of $11^\circ$, with an additional $11^\circ$ twist out of the plane. The mirror diameter is 3 mm to satisfy geometric constraints. The mode is circular at the position of the atoms with a waist of $w_0 = 10.0\,\mu$m. The round-trip length is $L = 6.96$\,cm, corresponding to a free spectral range of 4.3\,GHz and a splitting between the co-propagating circularly polarized modes of 133\,MHz. We note that planar ring cavities with a similar value of $w_0/\lambda$ have recently been demonstrated with $\mathcal{F} \approx 51,000$~\cite{chen2022b}.

\subsection{Twisted cavity}

The cavity has four resonance modes in the fundamental transverse mode (TEM$_{00}$) for each longitudinal mode number, corresponding to two polarization ($\sigma^+$ and $\sigma^-$), and clockwise (CW) and counter-clockwise (CCW) propagation directions. For a non-zero twist angle $\theta$, the polarization of light rotates in each round trip, which manifests as a phase shift with opposite sign for $\sigma^\pm$, splitting their resonance frequencies. However, the CW $\sigma^\pm$ mode is degenerate with the CCW $\sigma^\mp$ mode, as required by time-reversal symmetry. Therefore, if we align the atomic $\sigma^+$ and $\sigma^-$ transition frequencies with the two cavity frequencies, both transitions will decay via emission of photons in the same direction. Here, the time-reversal symmetry is broken by the magnetic field. If the direction of $\vec{B}$ is reversed, the emission direction will also reverse.

If the mirror coatings are birefringent, the eigenmodes are no longer perfectly circularly polarized. This effect can be minimized through the use of appropriately designed coatings, and very high levels of circular polarization have been demonstrated~\cite{jaffe2022a}. If the polarization is elliptical, some photon emission will go into modes propagating in the opposite direction. As these photons leave the cavity through a different port, this only affects the entanglement rate, but not the fidelity.

\subsection{Atom-cavity coupling strength}

In a Fabry-Perot cavity, the cooperativity for a two-level atom at the maximum of the electric field is given by $C_{fp} = 24 \mathcal{F}/(\pi k^2 w_0^2)$~\cite{tanji-suzuki2011}, where $\mathcal{F}$ is the cavity finesse $\mathcal{F} = \pi c/(L \kappa)$, $k = 2\pi/\lambda$, $w_0$ is the cavity waist, and $L$ is the mirror separation. In a running-wave cavity, the cooperativity is lower by a factor of 4, because of the absence of constructive interference between the forward and return beams~\cite{chen2022b}. Therefore, we have $C = 6 \mathcal{F}/(\pi k^2 w_0^2)$, but now with $\mathcal{F} = 2 \pi c/(L \kappa)$, where $L$ is the cavity round trip length. In a multi-level atom, this is further reduced by the branching ratio $R_{br}$ from the target excited state to the target ground state. We derive the atom-cavity coupling strength $g$ using the relation $C = 4 g^2/(\kappa \Gamma)$, where $\Gamma$ is the atomic transition decay rate (total decay rate, not corrected for $R_{br}$). This yields:

\begin{equation}
\label{eq:g}
    g = \sqrt{\frac{3 c R_{br} \Gamma}{k^2 w_0^2 L}}
\end{equation}

\section{Fidelity simulations}
\label{sec:sims}

\begin{figure}
    \centering
    \includegraphics[width=\columnwidth]{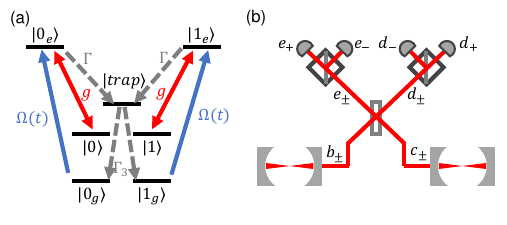}
    \caption{Fidelity simulation.
    (a) Simplified level diagram used in the fidelity simulation. 
    (b) Schematic of the photon measurement apparatus.
    }
    \label{fig:Appen_Fig2.png}
\end{figure}

To estimate the achievable entanglement rate and the contribution of various imperfections to the resulting state fidelity, we perform simulations of the atom-cavity system using the Lindblad master equation. We consider the simplified level diagram in Fig. ~\ref{fig:Appen_Fig2.png}a, described by the following Hamiltonian:

\begin{equation}
\begin{split}
    H =  & \Delta \cdot \left(|0_e\rangle\langle 0_e| + |1_e\rangle\langle 1_e|\right) \\
    &+ g\cdot \left(|0_e\rangle\langle 0|a_-+|0\rangle\langle 0_e|a_-^\dagger + |1_e\rangle\langle 1|a_+ +|1\rangle\langle 1_e|a_+^\dagger\right)\\
    &+ \Omega(t)/2 \cdot \left(|0_e\rangle\langle 0_g| + |0_g\rangle\langle 0_e| + |1_e\rangle\langle 1_g| + |1_g\rangle\langle 1_e|\right)
\end{split}
\end{equation}

Here, $\Delta$ is a detuning of the atomic transitions with respect to the cavity resonance (\emph{i.e.}, from Doppler shifts, or from a deliberate light shift), and $\Omega(t)$ is the Rabi frequency of the two-photon excitation pulse (the intermediate state is not included explicitly).

We include the following jump operators:

\begin{equation}
    \begin{aligned}
    c_1 &= \sqrt{\Gamma (1-R_{br})} |trap\rangle\langle 0_e|, \\
    c_2 &=  \sqrt{\Gamma(1-R_{br})} |trap\rangle\langle 1_e|, \\
    c_3 &= \sqrt{\Gamma R_{br}} |0\rangle\langle 0_e|,\\
    c_4 &=  \sqrt{\Gamma R_{br}} |1\rangle\langle 1_e|, \\
    c_5 &= \sqrt{\kappa} a_-, \\
    c_6 &= \sqrt{\kappa} a_+, \\
    c_7 &= \sqrt{\Gamma_3/2} |0_g\rangle\langle trap|, \\
    c_8 &= \sqrt{\Gamma_3/2} |1_g\rangle\langle trap|, 
    \end{aligned}
\end{equation}
where $|trap\rangle$ is a simplified representation of the $^3P_1$ manifold, and has equal probability of decaying into either ground state, $|0_g\rangle$ or $|1_g\rangle$. We take the cavity output field to be $a_{out,\pm} = \sqrt{\kappa} a_\pm$, following the input-output formalism~\cite{gardiner2004quantum}.

\subsection{Bell pair fidelity}

The successful generation of Bell pairs is heralded by coincident two-photon detection in the measurement apparatus shown in Fig. ~\ref{fig:Appen_Fig2.png}b. The output modes are related to the input modes using the beamsplitter relation~\cite{Kiraz2004}:

\begin{equation}
\left(
    \begin{array}{cc}
         d_\pm  \\
         e_\pm 
    \end{array}
\right) = 
\left(
    \begin{array}{cc}
         \cos \xi & -e^{-i \phi}\sin \xi  \\
         e^{i \phi} \sin \xi  & \cos \xi 
    \end{array}
\right)
\left(
    \begin{array}{cc}
         b_\pm  \\
         c_\pm 
    \end{array}
\right)
\end{equation}

A coincidence between $d_+$ and $e_-$ or $e_+$ and $d_-$ heralds the Bell state $\ket{\psi^-} = (\ket{10} - \ket{01})/\sqrt{2}$, while a coincidence between $d_+$ and $d_-$ or $e_+$ and $e_-$ heralds $\ket{\psi^+}$. Imperfections in the experiment give rise to a faulty atomic state, with an error probability that depends on the detection time of the photons. We determine this from simulation by evaluating the expectation values of the stabilizers of the Bell state as a function of the photon detection times, using the quantum regression theorem~\cite{gardiner2004quantum} as implemented in QuTIP~\cite{qutip}. 
We first consider Bell state $|\psi^-\rangle$, evaluating the following two-time correlation functions involving photon detection and the Pauli operators $X_i$, $Z_i$ acting on the qubit subspace $\{\ket{0},\ket{1}\}$ on the $i$th qubit:

\begin{align}
    P(t_1,t_2) &= \langle d_+^\dag(t_1) e_-^\dag(t_2) e_-(t_2) d_+(t_1) \rangle \\
    XX(t_1,t_2) &= \langle d_+^\dag(t_1) e_-^\dag(t_2) X_1(t_2) X_2(t_2) e_-(t_2) d_+(t_1) \rangle \\
    ZZ(t_1,t_2) &= \langle d_+^\dag(t_1) e_-^\dag(t_2) Z_1(t_2) Z_2(t_2) e_-(t_2) d_+(t_1) \rangle \\
    L(t_1,t_2) &= \langle d_+^\dag(t_1) e_-^\dag(t_2) Z_1^2(t_2) Z_2^2(t_2) e_-(t_2) d_+(t_1) \rangle
\end{align}

Here, $P(t_1,t_2)$ is the probability density for a coincidence event to occur at times $(t_1,t_2)$. $XX$ and $ZZ$ are the expectation values  of the same probability multiplied by the $X_1X_2$ and $Z_1 Z_2$ stabilizers, and $L$ is the detection probability multiplied by the population in the qubit subspace. The ideal value of the $X_1 X_2$ stabilizer is $-1$, which allows us to define a probability of remaining in the qubit subspace and having a $Z$ error on the Bell state, conditioned on detecting a photon at $(t_1,t_2)$:

\begin{equation}
\label{eq:pz_cond}
    p_z(t_1,t_2) = \frac{1}{2}\left[ 1+\frac{XX(t_1,t_2)}{P(t_1,t_2)}\left /\frac{L(t_1,t_2)}{P(t_1,t_2)} \right. \right].
\end{equation}

We can analogously define the probability of an $X$ error in the qubit subspace, $p_x$, and the probability of a leakage error, $p_l$, as:

\begin{align}
    p_x(t_1,t_2) &= \frac{1}{2}\left[1+\frac{ZZ(t_1,t_2)}{P(t_1,t_2)}\left /\frac{L(t_1,t_2)}{P(t_1,t_2)} \right.\right] \\
    \label{eq:pl_cond}
    p_l(t_1,t_2) &= \left[1 - \frac{L(t_1,t_2)}{P(t_1,t_2)}\right]
\end{align}

Finally, the total probability of each type of error is determined by the weighted average:

\begin{equation}
\label{eq:Pz_total}
    P_z = \frac{1}{\mathcal{N}} \iint p_z(t_1,t_2) P(t_1,t_2) dt_1 dt_2
\end{equation}
where $\mathcal{N} = \iint P(t_1,t_2) dt_1 dt_2$. An analogous definition follows for $P_x$, $P_l$, and the total error probability is the sum of all three types of errors. The plots in Fig.~\ref{fig:fig3} show $P_z + P_x + P_l$, while Eqs.~\eqref{eq:pz_cond}-\eqref{eq:pl_cond} are used to generate the plots in Fig.~\ref{fig:fig4}a-c.

Analogous expressions can be defined for the $\ket{\psi^+}$ Bell state with the opposite coincidence pattern in modes $d_+$, $d_-$. However, they are  the same up to expected sign of the $X_1 X_2$ stabilizer in Eq.~\eqref{eq:pz_cond}, so we do not consider this case separately.

\bibliography{jdt_new, yiyi}

\begin{thebibliography}{76}%
\makeatletter
\providecommand \@ifxundefined [1]{%
 \@ifx{#1\undefined}
}%
\providecommand \@ifnum [1]{%
 \ifnum #1\expandafter \@firstoftwo
 \else \expandafter \@secondoftwo
 \fi
}%
\providecommand \@ifx [1]{%
 \ifx #1\expandafter \@firstoftwo
 \else \expandafter \@secondoftwo
 \fi
}%
\providecommand \natexlab [1]{#1}%
\providecommand \enquote  [1]{``#1''}%
\providecommand \bibnamefont  [1]{#1}%
\providecommand \bibfnamefont [1]{#1}%
\providecommand \citenamefont [1]{#1}%
\providecommand \href@noop [0]{\@secondoftwo}%
\providecommand \href [0]{\begingroup \@sanitize@url \@href}%
\providecommand \@href[1]{\@@startlink{#1}\@@href}%
\providecommand \@@href[1]{\endgroup#1\@@endlink}%
\providecommand \@sanitize@url [0]{\catcode `\\12\catcode `\$12\catcode
  `\&12\catcode `\#12\catcode `\^12\catcode `\_12\catcode `\%12\relax}%
\providecommand \@@startlink[1]{}%
\providecommand \@@endlink[0]{}%
\providecommand \url  [0]{\begingroup\@sanitize@url \@url }%
\providecommand \@url [1]{\endgroup\@href {#1}{\urlprefix }}%
\providecommand \urlprefix  [0]{URL }%
\providecommand \Eprint [0]{\href }%
\providecommand \doibase [0]{https://doi.org/}%
\providecommand \selectlanguage [0]{\@gobble}%
\providecommand \bibinfo  [0]{\@secondoftwo}%
\providecommand \bibfield  [0]{\@secondoftwo}%
\providecommand \translation [1]{[#1]}%
\providecommand \BibitemOpen [0]{}%
\providecommand \bibitemStop [0]{}%
\providecommand \bibitemNoStop [0]{.\EOS\space}%
\providecommand \EOS [0]{\spacefactor3000\relax}%
\providecommand \BibitemShut  [1]{\csname bibitem#1\endcsname}%
\let\auto@bib@innerbib\@empty
%</preamble>
\bibitem [{\citenamefont {Gidney}\ and\ \citenamefont
  {Eker{\aa}}(2021)}]{gidney2021a}%
  \BibitemOpen
  \bibfield  {author} {\bibinfo {author} {\bibfnamefont {C.}~\bibnamefont
  {Gidney}}\ and\ \bibinfo {author} {\bibfnamefont {M.}~\bibnamefont
  {Eker{\aa}}},\ }\bibfield  {title} {\bibinfo {title} {How to factor 2048 bit
  {{RSA}} integers in 8 hours using 20 million noisy qubits},\ }\href@noop {}
  {\bibfield  {journal} {\bibinfo  {journal} {Quantum}\ }\textbf {\bibinfo
  {volume} {5}},\ \bibinfo {pages} {433} (\bibinfo {year} {2021})}\BibitemShut
  {NoStop}%
\bibitem [{\citenamefont {Beverland}\ \emph {et~al.}(2022)\citenamefont
  {Beverland}, \citenamefont {Murali}, \citenamefont {Troyer}, \citenamefont
  {Svore}, \citenamefont {Hoefler}, \citenamefont {Kliuchnikov}, \citenamefont
  {Low}, \citenamefont {Soeken}, \citenamefont {Sundaram},\ and\ \citenamefont
  {Vaschillo}}]{beverland2022}%
  \BibitemOpen
  \bibfield  {author} {\bibinfo {author} {\bibfnamefont {M.~E.}\ \bibnamefont
  {Beverland}}, \bibinfo {author} {\bibfnamefont {P.}~\bibnamefont {Murali}},
  \bibinfo {author} {\bibfnamefont {M.}~\bibnamefont {Troyer}}, \bibinfo
  {author} {\bibfnamefont {K.~M.}\ \bibnamefont {Svore}}, \bibinfo {author}
  {\bibfnamefont {T.}~\bibnamefont {Hoefler}}, \bibinfo {author} {\bibfnamefont
  {V.}~\bibnamefont {Kliuchnikov}}, \bibinfo {author} {\bibfnamefont {G.~H.}\
  \bibnamefont {Low}}, \bibinfo {author} {\bibfnamefont {M.}~\bibnamefont
  {Soeken}}, \bibinfo {author} {\bibfnamefont {A.}~\bibnamefont {Sundaram}},\
  and\ \bibinfo {author} {\bibfnamefont {A.}~\bibnamefont {Vaschillo}},\ }\href
  {http://arxiv.org/abs/2211.07629} {\bibinfo {title} {Assessing requirements
  to scale to practical quantum advantage}} (\bibinfo {year} {2022}),\ \Eprint
  {https://arxiv.org/abs/2211.07629} {arxiv:2211.07629 [quant-ph]} \BibitemShut
  {NoStop}%
\bibitem [{\citenamefont {Meter}\ \emph {et~al.}(2008)\citenamefont {Meter},
  \citenamefont {Munro}, \citenamefont {Nemoto},\ and\ \citenamefont
  {Itoh}}]{meter2008}%
  \BibitemOpen
  \bibfield  {author} {\bibinfo {author} {\bibfnamefont {R.~V.}\ \bibnamefont
  {Meter}}, \bibinfo {author} {\bibfnamefont {W.~J.}\ \bibnamefont {Munro}},
  \bibinfo {author} {\bibfnamefont {K.}~\bibnamefont {Nemoto}},\ and\ \bibinfo
  {author} {\bibfnamefont {K.~M.}\ \bibnamefont {Itoh}},\ }\bibfield  {title}
  {\bibinfo {title} {Arithmetic on a distributed-memory quantum
  multicomputer},\ }\href {https://doi.org/10.1145/1324177.1324179} {\bibfield
  {journal} {\bibinfo  {journal} {ACM Journal on Emerging Technologies in
  Computing Systems}\ }\textbf {\bibinfo {volume} {3}},\ \bibinfo {pages} {2:1}
  (\bibinfo {year} {2008})}\BibitemShut {NoStop}%
\bibitem [{\citenamefont {Monroe}\ \emph {et~al.}(2014)\citenamefont {Monroe},
  \citenamefont {Raussendorf}, \citenamefont {Ruthven}, \citenamefont {Brown},
  \citenamefont {Maunz}, \citenamefont {Duan},\ and\ \citenamefont
  {Kim}}]{Monroe2014}%
  \BibitemOpen
  \bibfield  {author} {\bibinfo {author} {\bibfnamefont {C.}~\bibnamefont
  {Monroe}}, \bibinfo {author} {\bibfnamefont {R.}~\bibnamefont {Raussendorf}},
  \bibinfo {author} {\bibfnamefont {A.}~\bibnamefont {Ruthven}}, \bibinfo
  {author} {\bibfnamefont {K.~R.}\ \bibnamefont {Brown}}, \bibinfo {author}
  {\bibfnamefont {P.}~\bibnamefont {Maunz}}, \bibinfo {author} {\bibfnamefont
  {L.~M.}\ \bibnamefont {Duan}},\ and\ \bibinfo {author} {\bibfnamefont
  {J.}~\bibnamefont {Kim}},\ }\bibfield  {title} {\bibinfo {title} {Large-scale
  modular quantum-computer architecture with atomic memory and photonic
  interconnects},\ }\href {http://link.aps.org/doi/10.1103/PhysRevA.89.022317}
  {\bibfield  {journal} {\bibinfo  {journal} {Physical Review A}\ }\textbf
  {\bibinfo {volume} {89}},\ \bibinfo {pages} {022317} (\bibinfo {year}
  {2014})}\BibitemShut {NoStop}%
\bibitem [{\citenamefont {Nickerson}\ \emph {et~al.}(2014)\citenamefont
  {Nickerson}, \citenamefont {Fitzsimons},\ and\ \citenamefont
  {Benjamin}}]{nickerson2014}%
  \BibitemOpen
  \bibfield  {author} {\bibinfo {author} {\bibfnamefont {N.~H.}\ \bibnamefont
  {Nickerson}}, \bibinfo {author} {\bibfnamefont {J.~F.}\ \bibnamefont
  {Fitzsimons}},\ and\ \bibinfo {author} {\bibfnamefont {S.~C.}\ \bibnamefont
  {Benjamin}},\ }\bibfield  {title} {\bibinfo {title} {Freely {{Scalable
  Quantum Technologies Using Cells}} of 5-to-50 {{Qubits}} with {{Very Lossy}}
  and {{Noisy Photonic Links}}},\ }\href
  {https://doi.org/10.1103/PhysRevX.4.041041} {\bibfield  {journal} {\bibinfo
  {journal} {Physical Review X}\ }\textbf {\bibinfo {volume} {4}},\ \bibinfo
  {pages} {041041} (\bibinfo {year} {2014})}\BibitemShut {NoStop}%
\bibitem [{\citenamefont {Li}\ and\ \citenamefont {Benjamin}(2016)}]{li2016}%
  \BibitemOpen
  \bibfield  {author} {\bibinfo {author} {\bibfnamefont {Y.}~\bibnamefont
  {Li}}\ and\ \bibinfo {author} {\bibfnamefont {S.~C.}\ \bibnamefont
  {Benjamin}},\ }\bibfield  {title} {\bibinfo {title} {Hierarchical surface
  code for network quantum computing with modules of arbitrary size},\ }\href
  {https://doi.org/10.1103/PhysRevA.94.042303} {\bibfield  {journal} {\bibinfo
  {journal} {Physical Review A}\ }\textbf {\bibinfo {volume} {94}},\ \bibinfo
  {pages} {042303} (\bibinfo {year} {2016})}\BibitemShut {NoStop}%
\bibitem [{\citenamefont {Gottesman}\ and\ \citenamefont
  {Chuang}(1999)}]{gottesman1999a}%
  \BibitemOpen
  \bibfield  {author} {\bibinfo {author} {\bibfnamefont {D.}~\bibnamefont
  {Gottesman}}\ and\ \bibinfo {author} {\bibfnamefont {I.~L.}\ \bibnamefont
  {Chuang}},\ }\bibfield  {title} {\bibinfo {title} {Demonstrating the
  viability of universal quantum computation using teleportation and
  single-qubit operations},\ }\href {https://doi.org/10.1038/46503} {\bibfield
  {journal} {\bibinfo  {journal} {Nature}\ }\textbf {\bibinfo {volume} {402}},\
  \bibinfo {pages} {390} (\bibinfo {year} {1999})}\BibitemShut {NoStop}%
\bibitem [{\citenamefont {Olmschenk}\ \emph {et~al.}(2009)\citenamefont
  {Olmschenk}, \citenamefont {Matsukevich}, \citenamefont {Maunz},
  \citenamefont {Hayes}, \citenamefont {Duan},\ and\ \citenamefont
  {Monroe}}]{Olmschenk2009}%
  \BibitemOpen
  \bibfield  {author} {\bibinfo {author} {\bibfnamefont {S.}~\bibnamefont
  {Olmschenk}}, \bibinfo {author} {\bibfnamefont {D.~N.}\ \bibnamefont
  {Matsukevich}}, \bibinfo {author} {\bibfnamefont {P.}~\bibnamefont {Maunz}},
  \bibinfo {author} {\bibfnamefont {D.}~\bibnamefont {Hayes}}, \bibinfo
  {author} {\bibfnamefont {L.-M.}\ \bibnamefont {Duan}},\ and\ \bibinfo
  {author} {\bibfnamefont {C.}~\bibnamefont {Monroe}},\ }\bibfield  {title}
  {\bibinfo {title} {Quantum teleportation between distant matter qubits},\
  }\href {https://doi.org/10.1126/science.1167209} {\bibfield  {journal}
  {\bibinfo  {journal} {Science}\ }\textbf {\bibinfo {volume} {323}},\ \bibinfo
  {pages} {486} (\bibinfo {year} {2009})}\BibitemShut {NoStop}%
\bibitem [{\citenamefont {Pfaff}\ \emph {et~al.}(2014)\citenamefont {Pfaff},
  \citenamefont {Hensen}, \citenamefont {Bernien}, \citenamefont {{van Dam}},
  \citenamefont {Blok}, \citenamefont {Taminiau}, \citenamefont {Tiggelman},
  \citenamefont {Schouten}, \citenamefont {Markham}, \citenamefont {Twitchen},\
  and\ \citenamefont {Hanson}}]{pfaff2014}%
  \BibitemOpen
  \bibfield  {author} {\bibinfo {author} {\bibfnamefont {W.}~\bibnamefont
  {Pfaff}}, \bibinfo {author} {\bibfnamefont {B.~J.}\ \bibnamefont {Hensen}},
  \bibinfo {author} {\bibfnamefont {H.}~\bibnamefont {Bernien}}, \bibinfo
  {author} {\bibfnamefont {S.~B.}\ \bibnamefont {{van Dam}}}, \bibinfo {author}
  {\bibfnamefont {M.~S.}\ \bibnamefont {Blok}}, \bibinfo {author}
  {\bibfnamefont {T.~H.}\ \bibnamefont {Taminiau}}, \bibinfo {author}
  {\bibfnamefont {M.~J.}\ \bibnamefont {Tiggelman}}, \bibinfo {author}
  {\bibfnamefont {R.~N.}\ \bibnamefont {Schouten}}, \bibinfo {author}
  {\bibfnamefont {M.}~\bibnamefont {Markham}}, \bibinfo {author} {\bibfnamefont
  {D.~J.}\ \bibnamefont {Twitchen}},\ and\ \bibinfo {author} {\bibfnamefont
  {R.}~\bibnamefont {Hanson}},\ }\bibfield  {title} {\bibinfo {title}
  {Unconditional quantum teleportation between distant solid-state quantum
  bits},\ }\href {http://www.sciencemag.org/content/345/6196/532.abstract}
  {\bibfield  {journal} {\bibinfo  {journal} {Science}\ }\textbf {\bibinfo
  {volume} {345}},\ \bibinfo {pages} {532} (\bibinfo {year}
  {2014})}\BibitemShut {NoStop}%
\bibitem [{\citenamefont {Chou}\ \emph {et~al.}(2018)\citenamefont {Chou},
  \citenamefont {Blumoff}, \citenamefont {Wang}, \citenamefont {Reinhold},
  \citenamefont {Axline}, \citenamefont {Gao}, \citenamefont {Frunzio},
  \citenamefont {Devoret}, \citenamefont {Jiang},\ and\ \citenamefont
  {Schoelkopf}}]{Chou2018}%
  \BibitemOpen
  \bibfield  {author} {\bibinfo {author} {\bibfnamefont {K.~S.}\ \bibnamefont
  {Chou}}, \bibinfo {author} {\bibfnamefont {J.~Z.}\ \bibnamefont {Blumoff}},
  \bibinfo {author} {\bibfnamefont {C.~S.}\ \bibnamefont {Wang}}, \bibinfo
  {author} {\bibfnamefont {P.~C.}\ \bibnamefont {Reinhold}}, \bibinfo {author}
  {\bibfnamefont {C.~J.}\ \bibnamefont {Axline}}, \bibinfo {author}
  {\bibfnamefont {Y.~Y.}\ \bibnamefont {Gao}}, \bibinfo {author} {\bibfnamefont
  {L.}~\bibnamefont {Frunzio}}, \bibinfo {author} {\bibfnamefont {M.~H.}\
  \bibnamefont {Devoret}}, \bibinfo {author} {\bibfnamefont {L.}~\bibnamefont
  {Jiang}},\ and\ \bibinfo {author} {\bibfnamefont {R.~J.}\ \bibnamefont
  {Schoelkopf}},\ }\bibfield  {title} {\bibinfo {title} {Deterministic
  teleportation of a quantum gate between two logical qubits},\ }\href
  {https://doi.org/10.1038/s41586-018-0470-y} {\bibfield  {journal} {\bibinfo
  {journal} {Nature}\ }\textbf {\bibinfo {volume} {561}},\ \bibinfo {pages}
  {368} (\bibinfo {year} {2018})}\BibitemShut {NoStop}%
\bibitem [{\citenamefont {Storz}\ \emph {et~al.}(2023)\citenamefont {Storz},
  \citenamefont {Sch{\"a}r}, \citenamefont {Kulikov}, \citenamefont {Magnard},
  \citenamefont {Kurpiers}, \citenamefont {L{\"u}tolf}, \citenamefont {Walter},
  \citenamefont {Copetudo}, \citenamefont {Reuer}, \citenamefont {Akin},
  \citenamefont {Besse}, \citenamefont {Gabureac}, \citenamefont {Norris},
  \citenamefont {Rosario}, \citenamefont {Martin}, \citenamefont {Martinez},
  \citenamefont {Amaya}, \citenamefont {Mitchell}, \citenamefont {Abellan},
  \citenamefont {Bancal}, \citenamefont {Sangouard}, \citenamefont {Royer},
  \citenamefont {Blais},\ and\ \citenamefont {Wallraff}}]{storz2023}%
  \BibitemOpen
  \bibfield  {author} {\bibinfo {author} {\bibfnamefont {S.}~\bibnamefont
  {Storz}}, \bibinfo {author} {\bibfnamefont {J.}~\bibnamefont {Sch{\"a}r}},
  \bibinfo {author} {\bibfnamefont {A.}~\bibnamefont {Kulikov}}, \bibinfo
  {author} {\bibfnamefont {P.}~\bibnamefont {Magnard}}, \bibinfo {author}
  {\bibfnamefont {P.}~\bibnamefont {Kurpiers}}, \bibinfo {author}
  {\bibfnamefont {J.}~\bibnamefont {L{\"u}tolf}}, \bibinfo {author}
  {\bibfnamefont {T.}~\bibnamefont {Walter}}, \bibinfo {author} {\bibfnamefont
  {A.}~\bibnamefont {Copetudo}}, \bibinfo {author} {\bibfnamefont
  {K.}~\bibnamefont {Reuer}}, \bibinfo {author} {\bibfnamefont
  {A.}~\bibnamefont {Akin}}, \bibinfo {author} {\bibfnamefont {J.-C.}\
  \bibnamefont {Besse}}, \bibinfo {author} {\bibfnamefont {M.}~\bibnamefont
  {Gabureac}}, \bibinfo {author} {\bibfnamefont {G.~J.}\ \bibnamefont
  {Norris}}, \bibinfo {author} {\bibfnamefont {A.}~\bibnamefont {Rosario}},
  \bibinfo {author} {\bibfnamefont {F.}~\bibnamefont {Martin}}, \bibinfo
  {author} {\bibfnamefont {J.}~\bibnamefont {Martinez}}, \bibinfo {author}
  {\bibfnamefont {W.}~\bibnamefont {Amaya}}, \bibinfo {author} {\bibfnamefont
  {M.~W.}\ \bibnamefont {Mitchell}}, \bibinfo {author} {\bibfnamefont
  {C.}~\bibnamefont {Abellan}}, \bibinfo {author} {\bibfnamefont {J.-D.}\
  \bibnamefont {Bancal}}, \bibinfo {author} {\bibfnamefont {N.}~\bibnamefont
  {Sangouard}}, \bibinfo {author} {\bibfnamefont {B.}~\bibnamefont {Royer}},
  \bibinfo {author} {\bibfnamefont {A.}~\bibnamefont {Blais}},\ and\ \bibinfo
  {author} {\bibfnamefont {A.}~\bibnamefont {Wallraff}},\ }\bibfield  {title}
  {\bibinfo {title} {Loophole-free {{Bell}} inequality violation with
  superconducting circuits},\ }\href
  {https://doi.org/10.1038/s41586-023-05885-0} {\bibfield  {journal} {\bibinfo
  {journal} {Nature}\ }\textbf {\bibinfo {volume} {617}},\ \bibinfo {pages}
  {265} (\bibinfo {year} {2023})}\BibitemShut {NoStop}%
\bibitem [{\citenamefont {Sahu}\ \emph {et~al.}(2023)\citenamefont {Sahu},
  \citenamefont {Qiu}, \citenamefont {Hease}, \citenamefont {Arnold},
  \citenamefont {Minoguchi}, \citenamefont {Rabl},\ and\ \citenamefont
  {Fink}}]{sahu2023}%
  \BibitemOpen
  \bibfield  {author} {\bibinfo {author} {\bibfnamefont {R.}~\bibnamefont
  {Sahu}}, \bibinfo {author} {\bibfnamefont {L.}~\bibnamefont {Qiu}}, \bibinfo
  {author} {\bibfnamefont {W.}~\bibnamefont {Hease}}, \bibinfo {author}
  {\bibfnamefont {G.}~\bibnamefont {Arnold}}, \bibinfo {author} {\bibfnamefont
  {Y.}~\bibnamefont {Minoguchi}}, \bibinfo {author} {\bibfnamefont
  {P.}~\bibnamefont {Rabl}},\ and\ \bibinfo {author} {\bibfnamefont {J.~M.}\
  \bibnamefont {Fink}},\ }\bibfield  {title} {\bibinfo {title} {Entangling
  microwaves with light},\ }\href {https://doi.org/10.1126/science.adg3812}
  {\bibfield  {journal} {\bibinfo  {journal} {Science}\ }\textbf {\bibinfo
  {volume} {380}},\ \bibinfo {pages} {718} (\bibinfo {year}
  {2023})}\BibitemShut {NoStop}%
\bibitem [{\citenamefont {Meesala}\ \emph {et~al.}(2023)\citenamefont
  {Meesala}, \citenamefont {Lake}, \citenamefont {Wood}, \citenamefont
  {Chiappina}, \citenamefont {Zhong}, \citenamefont {Beyer}, \citenamefont
  {Shaw}, \citenamefont {Jiang},\ and\ \citenamefont {Painter}}]{meesala2023}%
  \BibitemOpen
  \bibfield  {author} {\bibinfo {author} {\bibfnamefont {S.}~\bibnamefont
  {Meesala}}, \bibinfo {author} {\bibfnamefont {D.}~\bibnamefont {Lake}},
  \bibinfo {author} {\bibfnamefont {S.}~\bibnamefont {Wood}}, \bibinfo {author}
  {\bibfnamefont {P.}~\bibnamefont {Chiappina}}, \bibinfo {author}
  {\bibfnamefont {C.}~\bibnamefont {Zhong}}, \bibinfo {author} {\bibfnamefont
  {A.~D.}\ \bibnamefont {Beyer}}, \bibinfo {author} {\bibfnamefont {M.~D.}\
  \bibnamefont {Shaw}}, \bibinfo {author} {\bibfnamefont {L.}~\bibnamefont
  {Jiang}},\ and\ \bibinfo {author} {\bibfnamefont {O.}~\bibnamefont
  {Painter}},\ }\href {https://doi.org/10.48550/arXiv.2312.13559} {\bibinfo
  {title} {Quantum entanglement between optical and microwave photonic qubits}}
  (\bibinfo {year} {2023}),\ \Eprint {https://arxiv.org/abs/2312.13559}
  {arxiv:2312.13559 [quant-ph]} \BibitemShut {NoStop}%
\bibitem [{\citenamefont {Moehring}\ \emph
  {et~al.}(2007{\natexlab{a}})\citenamefont {Moehring}, \citenamefont {Maunz},
  \citenamefont {Olmschenk}, \citenamefont {Younge}, \citenamefont
  {Matsukevich}, \citenamefont {Duan},\ and\ \citenamefont
  {Monroe}}]{moehring2007}%
  \BibitemOpen
  \bibfield  {author} {\bibinfo {author} {\bibfnamefont {D.~L.}\ \bibnamefont
  {Moehring}}, \bibinfo {author} {\bibfnamefont {P.}~\bibnamefont {Maunz}},
  \bibinfo {author} {\bibfnamefont {S.}~\bibnamefont {Olmschenk}}, \bibinfo
  {author} {\bibfnamefont {K.~C.}\ \bibnamefont {Younge}}, \bibinfo {author}
  {\bibfnamefont {D.~N.}\ \bibnamefont {Matsukevich}}, \bibinfo {author}
  {\bibfnamefont {L.~M.}\ \bibnamefont {Duan}},\ and\ \bibinfo {author}
  {\bibfnamefont {C.}~\bibnamefont {Monroe}},\ }\bibfield  {title} {\bibinfo
  {title} {Entanglement of single-atom quantum bits at a distance},\ }\href
  {http://dx.doi.org/10.1038/nature06118} {\bibfield  {journal} {\bibinfo
  {journal} {Nature}\ }\textbf {\bibinfo {volume} {449}},\ \bibinfo {pages}
  {68} (\bibinfo {year} {2007}{\natexlab{a}})}\BibitemShut {NoStop}%
\bibitem [{\citenamefont {Ritter}\ \emph {et~al.}(2012)\citenamefont {Ritter},
  \citenamefont {N{\"o}lleke}, \citenamefont {Hahn}, \citenamefont {Reiserer},
  \citenamefont {Neuzner}, \citenamefont {Uphoff}, \citenamefont {M{\"u}cke},
  \citenamefont {Figueroa}, \citenamefont {Bochmann},\ and\ \citenamefont
  {Rempe}}]{ritter2012}%
  \BibitemOpen
  \bibfield  {author} {\bibinfo {author} {\bibfnamefont {S.}~\bibnamefont
  {Ritter}}, \bibinfo {author} {\bibfnamefont {C.}~\bibnamefont {N{\"o}lleke}},
  \bibinfo {author} {\bibfnamefont {C.}~\bibnamefont {Hahn}}, \bibinfo {author}
  {\bibfnamefont {A.}~\bibnamefont {Reiserer}}, \bibinfo {author}
  {\bibfnamefont {A.}~\bibnamefont {Neuzner}}, \bibinfo {author} {\bibfnamefont
  {M.}~\bibnamefont {Uphoff}}, \bibinfo {author} {\bibfnamefont
  {M.}~\bibnamefont {M{\"u}cke}}, \bibinfo {author} {\bibfnamefont
  {E.}~\bibnamefont {Figueroa}}, \bibinfo {author} {\bibfnamefont
  {J.}~\bibnamefont {Bochmann}},\ and\ \bibinfo {author} {\bibfnamefont
  {G.}~\bibnamefont {Rempe}},\ }\bibfield  {title} {\bibinfo {title} {An
  elementary quantum network of single atoms in optical cavities},\ }\href
  {http://www.nature.com/doifinder/10.1038/nature11023} {\bibfield  {journal}
  {\bibinfo  {journal} {Nature}\ }\textbf {\bibinfo {volume} {484}},\ \bibinfo
  {pages} {195} (\bibinfo {year} {2012})}\BibitemShut {NoStop}%
\bibitem [{\citenamefont {Hofmann}\ \emph {et~al.}(2012)\citenamefont
  {Hofmann}, \citenamefont {Krug}, \citenamefont {Ortegel}, \citenamefont
  {G{\'e}rard}, \citenamefont {Weber}, \citenamefont {Rosenfeld},\ and\
  \citenamefont {Weinfurter}}]{hofmann2012}%
  \BibitemOpen
  \bibfield  {author} {\bibinfo {author} {\bibfnamefont {J.}~\bibnamefont
  {Hofmann}}, \bibinfo {author} {\bibfnamefont {M.}~\bibnamefont {Krug}},
  \bibinfo {author} {\bibfnamefont {N.}~\bibnamefont {Ortegel}}, \bibinfo
  {author} {\bibfnamefont {L.}~\bibnamefont {G{\'e}rard}}, \bibinfo {author}
  {\bibfnamefont {M.}~\bibnamefont {Weber}}, \bibinfo {author} {\bibfnamefont
  {W.}~\bibnamefont {Rosenfeld}},\ and\ \bibinfo {author} {\bibfnamefont
  {H.}~\bibnamefont {Weinfurter}},\ }\bibfield  {title} {\bibinfo {title}
  {Heralded {{Entanglement Between Widely Separated Atoms}}},\ }\href
  {https://doi.org/10.1126/science.1221856} {\bibfield  {journal} {\bibinfo
  {journal} {Science}\ }\textbf {\bibinfo {volume} {337}},\ \bibinfo {pages}
  {72} (\bibinfo {year} {2012})}\BibitemShut {NoStop}%
\bibitem [{\citenamefont {Bernien}\ \emph {et~al.}(2013)\citenamefont
  {Bernien}, \citenamefont {Hensen}, \citenamefont {Pfaff}, \citenamefont
  {Koolstra}, \citenamefont {Blok}, \citenamefont {Robledo}, \citenamefont
  {Taminiau}, \citenamefont {Markham}, \citenamefont {Twitchen}, \citenamefont
  {Childress},\ and\ \citenamefont {Hanson}}]{bernien2013}%
  \BibitemOpen
  \bibfield  {author} {\bibinfo {author} {\bibfnamefont {H.}~\bibnamefont
  {Bernien}}, \bibinfo {author} {\bibfnamefont {B.}~\bibnamefont {Hensen}},
  \bibinfo {author} {\bibfnamefont {W.}~\bibnamefont {Pfaff}}, \bibinfo
  {author} {\bibfnamefont {G.}~\bibnamefont {Koolstra}}, \bibinfo {author}
  {\bibfnamefont {M.~S.}\ \bibnamefont {Blok}}, \bibinfo {author}
  {\bibfnamefont {L.}~\bibnamefont {Robledo}}, \bibinfo {author} {\bibfnamefont
  {T.~H.}\ \bibnamefont {Taminiau}}, \bibinfo {author} {\bibfnamefont
  {M.}~\bibnamefont {Markham}}, \bibinfo {author} {\bibfnamefont {D.~J.}\
  \bibnamefont {Twitchen}}, \bibinfo {author} {\bibfnamefont {L.}~\bibnamefont
  {Childress}},\ and\ \bibinfo {author} {\bibfnamefont {R.}~\bibnamefont
  {Hanson}},\ }\bibfield  {title} {\bibinfo {title} {Heralded entanglement
  between solid-state qubits separated by three metres},\ }\href
  {http://www.nature.com/doifinder/10.1038/nature12016} {\bibfield  {journal}
  {\bibinfo  {journal} {Nature}\ }\textbf {\bibinfo {volume} {497}},\ \bibinfo
  {pages} {86} (\bibinfo {year} {2013})}\BibitemShut {NoStop}%
\bibitem [{\citenamefont {Hucul}\ \emph {et~al.}(2014)\citenamefont {Hucul},
  \citenamefont {Inlek}, \citenamefont {Vittorini}, \citenamefont {Crocker},
  \citenamefont {Debnath}, \citenamefont {Clark},\ and\ \citenamefont
  {Monroe}}]{Hucul2014}%
  \BibitemOpen
  \bibfield  {author} {\bibinfo {author} {\bibfnamefont {D.}~\bibnamefont
  {Hucul}}, \bibinfo {author} {\bibfnamefont {I.~V.}\ \bibnamefont {Inlek}},
  \bibinfo {author} {\bibfnamefont {G.}~\bibnamefont {Vittorini}}, \bibinfo
  {author} {\bibfnamefont {C.}~\bibnamefont {Crocker}}, \bibinfo {author}
  {\bibfnamefont {S.}~\bibnamefont {Debnath}}, \bibinfo {author} {\bibfnamefont
  {S.~M.}\ \bibnamefont {Clark}},\ and\ \bibinfo {author} {\bibfnamefont
  {C.}~\bibnamefont {Monroe}},\ }\bibfield  {title} {\bibinfo {title} {Modular
  entanglement of atomic qubits using photons and phonons},\ }\href
  {http://www.nature.com/doifinder/10.1038/nphys3150} {\bibfield  {journal}
  {\bibinfo  {journal} {Nature Physics}\ }\textbf {\bibinfo {volume} {11}},\
  \bibinfo {pages} {37} (\bibinfo {year} {2014})}\BibitemShut {NoStop}%
\bibitem [{\citenamefont {Delteil}\ \emph {et~al.}(2016)\citenamefont
  {Delteil}, \citenamefont {Sun}, \citenamefont {Gao}, \citenamefont {Togan},
  \citenamefont {Faelt},\ and\ \citenamefont {Imamo{\u g}lu}}]{delteil2016}%
  \BibitemOpen
  \bibfield  {author} {\bibinfo {author} {\bibfnamefont {A.}~\bibnamefont
  {Delteil}}, \bibinfo {author} {\bibfnamefont {Z.}~\bibnamefont {Sun}},
  \bibinfo {author} {\bibfnamefont {W.-b.}\ \bibnamefont {Gao}}, \bibinfo
  {author} {\bibfnamefont {E.}~\bibnamefont {Togan}}, \bibinfo {author}
  {\bibfnamefont {S.}~\bibnamefont {Faelt}},\ and\ \bibinfo {author}
  {\bibfnamefont {A.}~\bibnamefont {Imamo{\u g}lu}},\ }\bibfield  {title}
  {\bibinfo {title} {Generation of heralded entanglement between distant hole
  spins},\ }\href {https://doi.org/10.1038/nphys3605} {\bibfield  {journal}
  {\bibinfo  {journal} {Nature Physics}\ }\textbf {\bibinfo {volume} {12}},\
  \bibinfo {pages} {218} (\bibinfo {year} {2016})}\BibitemShut {NoStop}%
\bibitem [{\citenamefont {Stephenson}\ \emph {et~al.}(2020)\citenamefont
  {Stephenson}, \citenamefont {Nadlinger}, \citenamefont {Nichol},
  \citenamefont {An}, \citenamefont {Drmota}, \citenamefont {Ballance},
  \citenamefont {Thirumalai}, \citenamefont {Goodwin}, \citenamefont {Lucas},\
  and\ \citenamefont {Ballance}}]{stephenson2020}%
  \BibitemOpen
  \bibfield  {author} {\bibinfo {author} {\bibfnamefont {L.~J.}\ \bibnamefont
  {Stephenson}}, \bibinfo {author} {\bibfnamefont {D.~P.}\ \bibnamefont
  {Nadlinger}}, \bibinfo {author} {\bibfnamefont {B.~C.}\ \bibnamefont
  {Nichol}}, \bibinfo {author} {\bibfnamefont {S.}~\bibnamefont {An}}, \bibinfo
  {author} {\bibfnamefont {P.}~\bibnamefont {Drmota}}, \bibinfo {author}
  {\bibfnamefont {T.~G.}\ \bibnamefont {Ballance}}, \bibinfo {author}
  {\bibfnamefont {K.}~\bibnamefont {Thirumalai}}, \bibinfo {author}
  {\bibfnamefont {J.~F.}\ \bibnamefont {Goodwin}}, \bibinfo {author}
  {\bibfnamefont {D.~M.}\ \bibnamefont {Lucas}},\ and\ \bibinfo {author}
  {\bibfnamefont {C.~J.}\ \bibnamefont {Ballance}},\ }\bibfield  {title}
  {\bibinfo {title} {High-{{Rate}}, {{High-Fidelity Entanglement}} of {{Qubits
  Across}} an {{Elementary Quantum Network}}},\ }\href
  {https://doi.org/10.1103/PhysRevLett.124.110501} {\bibfield  {journal}
  {\bibinfo  {journal} {Physical Review Letters}\ }\textbf {\bibinfo {volume}
  {124}},\ \bibinfo {pages} {110501} (\bibinfo {year} {2020})}\BibitemShut
  {NoStop}%
\bibitem [{\citenamefont {{van Leent}}\ \emph {et~al.}(2022)\citenamefont {{van
  Leent}}, \citenamefont {Bock}, \citenamefont {Fertig}, \citenamefont
  {Garthoff}, \citenamefont {Eppelt}, \citenamefont {Zhou}, \citenamefont
  {Malik}, \citenamefont {Seubert}, \citenamefont {Bauer}, \citenamefont
  {Rosenfeld}, \citenamefont {Zhang}, \citenamefont {Becher},\ and\
  \citenamefont {Weinfurter}}]{vanleent2022}%
  \BibitemOpen
  \bibfield  {author} {\bibinfo {author} {\bibfnamefont {T.}~\bibnamefont {{van
  Leent}}}, \bibinfo {author} {\bibfnamefont {M.}~\bibnamefont {Bock}},
  \bibinfo {author} {\bibfnamefont {F.}~\bibnamefont {Fertig}}, \bibinfo
  {author} {\bibfnamefont {R.}~\bibnamefont {Garthoff}}, \bibinfo {author}
  {\bibfnamefont {S.}~\bibnamefont {Eppelt}}, \bibinfo {author} {\bibfnamefont
  {Y.}~\bibnamefont {Zhou}}, \bibinfo {author} {\bibfnamefont {P.}~\bibnamefont
  {Malik}}, \bibinfo {author} {\bibfnamefont {M.}~\bibnamefont {Seubert}},
  \bibinfo {author} {\bibfnamefont {T.}~\bibnamefont {Bauer}}, \bibinfo
  {author} {\bibfnamefont {W.}~\bibnamefont {Rosenfeld}}, \bibinfo {author}
  {\bibfnamefont {W.}~\bibnamefont {Zhang}}, \bibinfo {author} {\bibfnamefont
  {C.}~\bibnamefont {Becher}},\ and\ \bibinfo {author} {\bibfnamefont
  {H.}~\bibnamefont {Weinfurter}},\ }\bibfield  {title} {\bibinfo {title}
  {Entangling single atoms over 33 km telecom fibre},\ }\href
  {https://doi.org/10.1038/s41586-022-04764-4} {\bibfield  {journal} {\bibinfo
  {journal} {Nature}\ }\textbf {\bibinfo {volume} {607}},\ \bibinfo {pages}
  {69} (\bibinfo {year} {2022})}\BibitemShut {NoStop}%
\bibitem [{\citenamefont {Krutyanskiy}\ \emph {et~al.}(2023)\citenamefont
  {Krutyanskiy}, \citenamefont {Galli}, \citenamefont {Krcmarsky},
  \citenamefont {Baier}, \citenamefont {Fioretto}, \citenamefont {Pu},
  \citenamefont {Mazloom}, \citenamefont {Sekatski}, \citenamefont {Canteri},
  \citenamefont {Teller}, \citenamefont {Schupp}, \citenamefont {Bate},
  \citenamefont {Meraner}, \citenamefont {Sangouard}, \citenamefont {Lanyon},\
  and\ \citenamefont {Northup}}]{krutyanskiy2023}%
  \BibitemOpen
  \bibfield  {author} {\bibinfo {author} {\bibfnamefont {V.}~\bibnamefont
  {Krutyanskiy}}, \bibinfo {author} {\bibfnamefont {M.}~\bibnamefont {Galli}},
  \bibinfo {author} {\bibfnamefont {V.}~\bibnamefont {Krcmarsky}}, \bibinfo
  {author} {\bibfnamefont {S.}~\bibnamefont {Baier}}, \bibinfo {author}
  {\bibfnamefont {D.~A.}\ \bibnamefont {Fioretto}}, \bibinfo {author}
  {\bibfnamefont {Y.}~\bibnamefont {Pu}}, \bibinfo {author} {\bibfnamefont
  {A.}~\bibnamefont {Mazloom}}, \bibinfo {author} {\bibfnamefont
  {P.}~\bibnamefont {Sekatski}}, \bibinfo {author} {\bibfnamefont
  {M.}~\bibnamefont {Canteri}}, \bibinfo {author} {\bibfnamefont
  {M.}~\bibnamefont {Teller}}, \bibinfo {author} {\bibfnamefont
  {J.}~\bibnamefont {Schupp}}, \bibinfo {author} {\bibfnamefont
  {J.}~\bibnamefont {Bate}}, \bibinfo {author} {\bibfnamefont {M.}~\bibnamefont
  {Meraner}}, \bibinfo {author} {\bibfnamefont {N.}~\bibnamefont {Sangouard}},
  \bibinfo {author} {\bibfnamefont {B.~P.}\ \bibnamefont {Lanyon}},\ and\
  \bibinfo {author} {\bibfnamefont {T.~E.}\ \bibnamefont {Northup}},\
  }\bibfield  {title} {\bibinfo {title} {Entanglement of {{Trapped-Ion Qubits
  Separated}} by 230 {{Meters}}},\ }\href
  {https://doi.org/10.1103/PhysRevLett.130.050803} {\bibfield  {journal}
  {\bibinfo  {journal} {Physical Review Letters}\ }\textbf {\bibinfo {volume}
  {130}},\ \bibinfo {pages} {050803} (\bibinfo {year} {2023})}\BibitemShut
  {NoStop}%
\bibitem [{\citenamefont {Knaut}\ \emph {et~al.}(2023)\citenamefont {Knaut},
  \citenamefont {Suleymanzade}, \citenamefont {Wei}, \citenamefont {Assumpcao},
  \citenamefont {Stas}, \citenamefont {Huan}, \citenamefont {Machielse},
  \citenamefont {Knall}, \citenamefont {Sutula}, \citenamefont {Baranes},
  \citenamefont {Sinclair}, \citenamefont {{De-Eknamkul}}, \citenamefont
  {Levonian}, \citenamefont {Bhaskar}, \citenamefont {Park}, \citenamefont
  {Lon{\v c}ar},\ and\ \citenamefont {Lukin}}]{knaut2023}%
  \BibitemOpen
  \bibfield  {author} {\bibinfo {author} {\bibfnamefont {C.~M.}\ \bibnamefont
  {Knaut}}, \bibinfo {author} {\bibfnamefont {A.}~\bibnamefont {Suleymanzade}},
  \bibinfo {author} {\bibfnamefont {Y.-C.}\ \bibnamefont {Wei}}, \bibinfo
  {author} {\bibfnamefont {D.~R.}\ \bibnamefont {Assumpcao}}, \bibinfo {author}
  {\bibfnamefont {P.-J.}\ \bibnamefont {Stas}}, \bibinfo {author}
  {\bibfnamefont {Y.~Q.}\ \bibnamefont {Huan}}, \bibinfo {author}
  {\bibfnamefont {B.}~\bibnamefont {Machielse}}, \bibinfo {author}
  {\bibfnamefont {E.~N.}\ \bibnamefont {Knall}}, \bibinfo {author}
  {\bibfnamefont {M.}~\bibnamefont {Sutula}}, \bibinfo {author} {\bibfnamefont
  {G.}~\bibnamefont {Baranes}}, \bibinfo {author} {\bibfnamefont
  {N.}~\bibnamefont {Sinclair}}, \bibinfo {author} {\bibfnamefont
  {C.}~\bibnamefont {{De-Eknamkul}}}, \bibinfo {author} {\bibfnamefont {D.~S.}\
  \bibnamefont {Levonian}}, \bibinfo {author} {\bibfnamefont {M.~K.}\
  \bibnamefont {Bhaskar}}, \bibinfo {author} {\bibfnamefont {H.}~\bibnamefont
  {Park}}, \bibinfo {author} {\bibfnamefont {M.}~\bibnamefont {Lon{\v c}ar}},\
  and\ \bibinfo {author} {\bibfnamefont {M.~D.}\ \bibnamefont {Lukin}},\ }\href
  {https://doi.org/10.48550/arXiv.2310.01316} {\bibinfo {title} {Entanglement
  of {{Nanophotonic Quantum Memory Nodes}} in a {{Telecommunication Network}}}}
  (\bibinfo {year} {2023}),\ \Eprint {https://arxiv.org/abs/2310.01316}
  {arxiv:2310.01316 [quant-ph]} \BibitemShut {NoStop}%
\bibitem [{\citenamefont {Bluvstein}\ \emph {et~al.}(2023)\citenamefont
  {Bluvstein}, \citenamefont {Evered}, \citenamefont {Geim}, \citenamefont
  {Li}, \citenamefont {Zhou}, \citenamefont {Manovitz}, \citenamefont {Ebadi},
  \citenamefont {Cain}, \citenamefont {Kalinowski}, \citenamefont {Hangleiter},
  \citenamefont {Ataides}, \citenamefont {Maskara}, \citenamefont {Cong},
  \citenamefont {Gao}, \citenamefont {Rodriguez}, \citenamefont {Karolyshyn},
  \citenamefont {Semeghini}, \citenamefont {Gullans}, \citenamefont {Greiner},
  \citenamefont {Vuletić},\ and\ \citenamefont {Lukin}}]{bluvstein2023}%
  \BibitemOpen
  \bibfield  {author} {\bibinfo {author} {\bibfnamefont {D.}~\bibnamefont
  {Bluvstein}}, \bibinfo {author} {\bibfnamefont {S.~J.}\ \bibnamefont
  {Evered}}, \bibinfo {author} {\bibfnamefont {A.~A.}\ \bibnamefont {Geim}},
  \bibinfo {author} {\bibfnamefont {S.~H.}\ \bibnamefont {Li}}, \bibinfo
  {author} {\bibfnamefont {H.}~\bibnamefont {Zhou}}, \bibinfo {author}
  {\bibfnamefont {T.}~\bibnamefont {Manovitz}}, \bibinfo {author}
  {\bibfnamefont {S.}~\bibnamefont {Ebadi}}, \bibinfo {author} {\bibfnamefont
  {M.}~\bibnamefont {Cain}}, \bibinfo {author} {\bibfnamefont {M.}~\bibnamefont
  {Kalinowski}}, \bibinfo {author} {\bibfnamefont {D.}~\bibnamefont
  {Hangleiter}}, \bibinfo {author} {\bibfnamefont {J.~P.~B.}\ \bibnamefont
  {Ataides}}, \bibinfo {author} {\bibfnamefont {N.}~\bibnamefont {Maskara}},
  \bibinfo {author} {\bibfnamefont {I.}~\bibnamefont {Cong}}, \bibinfo {author}
  {\bibfnamefont {X.}~\bibnamefont {Gao}}, \bibinfo {author} {\bibfnamefont
  {P.~S.}\ \bibnamefont {Rodriguez}}, \bibinfo {author} {\bibfnamefont
  {T.}~\bibnamefont {Karolyshyn}}, \bibinfo {author} {\bibfnamefont
  {G.}~\bibnamefont {Semeghini}}, \bibinfo {author} {\bibfnamefont {M.~J.}\
  \bibnamefont {Gullans}}, \bibinfo {author} {\bibfnamefont {M.}~\bibnamefont
  {Greiner}}, \bibinfo {author} {\bibfnamefont {V.}~\bibnamefont {Vuletić}},\
  and\ \bibinfo {author} {\bibfnamefont {M.~D.}\ \bibnamefont {Lukin}},\
  }\bibfield  {title} {\bibinfo {title} {Logical quantum processor based on
  reconfigurable atom arrays},\ }\href@noop {} {\bibfield  {journal} {\bibinfo
  {journal} {Nature}\ } (\bibinfo {year} {2023})}\BibitemShut {NoStop}%
\bibitem [{\citenamefont {Graham}\ \emph {et~al.}(2022)\citenamefont {Graham},
  \citenamefont {Song}, \citenamefont {Scott}, \citenamefont {Poole},
  \citenamefont {Phuttitarn}, \citenamefont {Jooya}, \citenamefont {Eichler},
  \citenamefont {Jiang}, \citenamefont {Marra}, \citenamefont {Grinkemeyer},
  \citenamefont {Kwon}, \citenamefont {Ebert}, \citenamefont {Cherek},
  \citenamefont {Lichtman}, \citenamefont {Gillette}, \citenamefont {Gilbert},
  \citenamefont {Bowman}, \citenamefont {Ballance}, \citenamefont {Campbell},
  \citenamefont {Dahl}, \citenamefont {Crawford}, \citenamefont {Blunt},
  \citenamefont {Rogers}, \citenamefont {Noel},\ and\ \citenamefont
  {Saffman}}]{graham2022a}%
  \BibitemOpen
  \bibfield  {author} {\bibinfo {author} {\bibfnamefont {T.~M.}\ \bibnamefont
  {Graham}}, \bibinfo {author} {\bibfnamefont {Y.}~\bibnamefont {Song}},
  \bibinfo {author} {\bibfnamefont {J.}~\bibnamefont {Scott}}, \bibinfo
  {author} {\bibfnamefont {C.}~\bibnamefont {Poole}}, \bibinfo {author}
  {\bibfnamefont {L.}~\bibnamefont {Phuttitarn}}, \bibinfo {author}
  {\bibfnamefont {K.}~\bibnamefont {Jooya}}, \bibinfo {author} {\bibfnamefont
  {P.}~\bibnamefont {Eichler}}, \bibinfo {author} {\bibfnamefont
  {X.}~\bibnamefont {Jiang}}, \bibinfo {author} {\bibfnamefont
  {A.}~\bibnamefont {Marra}}, \bibinfo {author} {\bibfnamefont
  {B.}~\bibnamefont {Grinkemeyer}}, \bibinfo {author} {\bibfnamefont
  {M.}~\bibnamefont {Kwon}}, \bibinfo {author} {\bibfnamefont {M.}~\bibnamefont
  {Ebert}}, \bibinfo {author} {\bibfnamefont {J.}~\bibnamefont {Cherek}},
  \bibinfo {author} {\bibfnamefont {M.~T.}\ \bibnamefont {Lichtman}}, \bibinfo
  {author} {\bibfnamefont {M.}~\bibnamefont {Gillette}}, \bibinfo {author}
  {\bibfnamefont {J.}~\bibnamefont {Gilbert}}, \bibinfo {author} {\bibfnamefont
  {D.}~\bibnamefont {Bowman}}, \bibinfo {author} {\bibfnamefont
  {T.}~\bibnamefont {Ballance}}, \bibinfo {author} {\bibfnamefont
  {C.}~\bibnamefont {Campbell}}, \bibinfo {author} {\bibfnamefont {E.~D.}\
  \bibnamefont {Dahl}}, \bibinfo {author} {\bibfnamefont {O.}~\bibnamefont
  {Crawford}}, \bibinfo {author} {\bibfnamefont {N.~S.}\ \bibnamefont {Blunt}},
  \bibinfo {author} {\bibfnamefont {B.}~\bibnamefont {Rogers}}, \bibinfo
  {author} {\bibfnamefont {T.}~\bibnamefont {Noel}},\ and\ \bibinfo {author}
  {\bibfnamefont {M.}~\bibnamefont {Saffman}},\ }\bibfield  {title} {\bibinfo
  {title} {Multi-qubit entanglement and algorithms on a neutral-atom quantum
  computer},\ }\href {https://doi.org/10.1038/s41586-022-04603-6} {\bibfield
  {journal} {\bibinfo  {journal} {Nature}\ }\textbf {\bibinfo {volume} {604}},\
  \bibinfo {pages} {457} (\bibinfo {year} {2022})}\BibitemShut {NoStop}%
\bibitem [{\citenamefont {Ma}\ \emph {et~al.}(2023)\citenamefont {Ma},
  \citenamefont {Liu}, \citenamefont {Peng}, \citenamefont {Zhang},
  \citenamefont {Jandura}, \citenamefont {Claes}, \citenamefont {Burgers},
  \citenamefont {Pupillo}, \citenamefont {Puri},\ and\ \citenamefont
  {Thompson}}]{ma2023a}%
  \BibitemOpen
  \bibfield  {author} {\bibinfo {author} {\bibfnamefont {S.}~\bibnamefont
  {Ma}}, \bibinfo {author} {\bibfnamefont {G.}~\bibnamefont {Liu}}, \bibinfo
  {author} {\bibfnamefont {P.}~\bibnamefont {Peng}}, \bibinfo {author}
  {\bibfnamefont {B.}~\bibnamefont {Zhang}}, \bibinfo {author} {\bibfnamefont
  {S.}~\bibnamefont {Jandura}}, \bibinfo {author} {\bibfnamefont
  {J.}~\bibnamefont {Claes}}, \bibinfo {author} {\bibfnamefont {A.~P.}\
  \bibnamefont {Burgers}}, \bibinfo {author} {\bibfnamefont {G.}~\bibnamefont
  {Pupillo}}, \bibinfo {author} {\bibfnamefont {S.}~\bibnamefont {Puri}},\ and\
  \bibinfo {author} {\bibfnamefont {J.~D.}\ \bibnamefont {Thompson}},\
  }\bibfield  {title} {\bibinfo {title} {High-fidelity gates and mid-circuit
  erasure conversion in an atomic qubit},\ }\href
  {https://doi.org/10.1038/s41586-023-06438-1} {\bibfield  {journal} {\bibinfo
  {journal} {Nature}\ }\textbf {\bibinfo {volume} {622}},\ \bibinfo {pages}
  {279} (\bibinfo {year} {2023})}\BibitemShut {NoStop}%
\bibitem [{\citenamefont {Huie}\ \emph {et~al.}(2023)\citenamefont {Huie},
  \citenamefont {Li}, \citenamefont {Chen}, \citenamefont {Hu}, \citenamefont
  {Jia}, \citenamefont {Sun},\ and\ \citenamefont {Covey}}]{huie2023a}%
  \BibitemOpen
  \bibfield  {author} {\bibinfo {author} {\bibfnamefont {W.}~\bibnamefont
  {Huie}}, \bibinfo {author} {\bibfnamefont {L.}~\bibnamefont {Li}}, \bibinfo
  {author} {\bibfnamefont {N.}~\bibnamefont {Chen}}, \bibinfo {author}
  {\bibfnamefont {X.}~\bibnamefont {Hu}}, \bibinfo {author} {\bibfnamefont
  {Z.}~\bibnamefont {Jia}}, \bibinfo {author} {\bibfnamefont {W.~K.~C.}\
  \bibnamefont {Sun}},\ and\ \bibinfo {author} {\bibfnamefont {J.~P.}\
  \bibnamefont {Covey}},\ }\bibfield  {title} {\bibinfo {title} {Repetitive
  readout and real-time control of nuclear spin qubits in ${}^{171}\mathrm{Yb}$
  atoms},\ }\href {https://doi.org/10.1103/PRXQuantum.4.030337} {\bibfield
  {journal} {\bibinfo  {journal} {PRX Quantum}\ }\textbf {\bibinfo {volume}
  {4}},\ \bibinfo {pages} {030337} (\bibinfo {year} {2023})}\BibitemShut
  {NoStop}%
\bibitem [{\citenamefont {Lis}\ \emph {et~al.}(2023)\citenamefont {Lis},
  \citenamefont {Senoo}, \citenamefont {McGrew}, \citenamefont {R{\"o}nchen},
  \citenamefont {Jenkins},\ and\ \citenamefont {Kaufman}}]{lis2023a}%
  \BibitemOpen
  \bibfield  {author} {\bibinfo {author} {\bibfnamefont {J.~W.}\ \bibnamefont
  {Lis}}, \bibinfo {author} {\bibfnamefont {A.}~\bibnamefont {Senoo}}, \bibinfo
  {author} {\bibfnamefont {W.~F.}\ \bibnamefont {McGrew}}, \bibinfo {author}
  {\bibfnamefont {F.}~\bibnamefont {R{\"o}nchen}}, \bibinfo {author}
  {\bibfnamefont {A.}~\bibnamefont {Jenkins}},\ and\ \bibinfo {author}
  {\bibfnamefont {A.~M.}\ \bibnamefont {Kaufman}},\ }\bibfield  {title}
  {\bibinfo {title} {Midcircuit {{Operations Using}} the {\emph{omg}}
  {{Architecture}} in {{Neutral Atom Arrays}}},\ }\href
  {https://doi.org/10.1103/PhysRevX.13.041035} {\bibfield  {journal} {\bibinfo
  {journal} {Physical Review X}\ }\textbf {\bibinfo {volume} {13}},\ \bibinfo
  {pages} {041035} (\bibinfo {year} {2023})}\BibitemShut {NoStop}%
\bibitem [{\citenamefont {Norcia}\ \emph {et~al.}(2023)\citenamefont {Norcia},
  \citenamefont {Cairncross}, \citenamefont {Barnes}, \citenamefont
  {Battaglino}, \citenamefont {Brown}, \citenamefont {Brown}, \citenamefont
  {Cassella}, \citenamefont {Chen}, \citenamefont {Coxe}, \citenamefont {Crow},
  \citenamefont {Epstein}, \citenamefont {Griger}, \citenamefont {Jones},
  \citenamefont {Kim}, \citenamefont {Kindem}, \citenamefont {King},
  \citenamefont {Kondov}, \citenamefont {Kotru}, \citenamefont {Lauigan},
  \citenamefont {Li}, \citenamefont {Lu}, \citenamefont {Megidish},
  \citenamefont {Marjanovic}, \citenamefont {McDonald}, \citenamefont
  {Mittiga}, \citenamefont {Muniz}, \citenamefont {Narayanaswami},
  \citenamefont {Nishiguchi}, \citenamefont {Notermans}, \citenamefont {Paule},
  \citenamefont {Pawlak}, \citenamefont {Peng}, \citenamefont {Ryou},
  \citenamefont {Smull}, \citenamefont {Stack}, \citenamefont {Stone},
  \citenamefont {Sucich}, \citenamefont {Urbanek}, \citenamefont {Van
  De~Veerdonk}, \citenamefont {Vendeiro}, \citenamefont {Wilkason},
  \citenamefont {Wu}, \citenamefont {Xie}, \citenamefont {Zhang},\ and\
  \citenamefont {Bloom}}]{norcia2023a}%
  \BibitemOpen
  \bibfield  {author} {\bibinfo {author} {\bibfnamefont {M.~A.}\ \bibnamefont
  {Norcia}}, \bibinfo {author} {\bibfnamefont {W.~B.}\ \bibnamefont
  {Cairncross}}, \bibinfo {author} {\bibfnamefont {K.}~\bibnamefont {Barnes}},
  \bibinfo {author} {\bibfnamefont {P.}~\bibnamefont {Battaglino}}, \bibinfo
  {author} {\bibfnamefont {A.}~\bibnamefont {Brown}}, \bibinfo {author}
  {\bibfnamefont {M.~O.}\ \bibnamefont {Brown}}, \bibinfo {author}
  {\bibfnamefont {K.}~\bibnamefont {Cassella}}, \bibinfo {author}
  {\bibfnamefont {C.-A.}\ \bibnamefont {Chen}}, \bibinfo {author}
  {\bibfnamefont {R.}~\bibnamefont {Coxe}}, \bibinfo {author} {\bibfnamefont
  {D.}~\bibnamefont {Crow}}, \bibinfo {author} {\bibfnamefont {J.}~\bibnamefont
  {Epstein}}, \bibinfo {author} {\bibfnamefont {C.}~\bibnamefont {Griger}},
  \bibinfo {author} {\bibfnamefont {A.~M.~W.}\ \bibnamefont {Jones}}, \bibinfo
  {author} {\bibfnamefont {H.}~\bibnamefont {Kim}}, \bibinfo {author}
  {\bibfnamefont {J.~M.}\ \bibnamefont {Kindem}}, \bibinfo {author}
  {\bibfnamefont {J.}~\bibnamefont {King}}, \bibinfo {author} {\bibfnamefont
  {S.~S.}\ \bibnamefont {Kondov}}, \bibinfo {author} {\bibfnamefont
  {K.}~\bibnamefont {Kotru}}, \bibinfo {author} {\bibfnamefont
  {J.}~\bibnamefont {Lauigan}}, \bibinfo {author} {\bibfnamefont
  {M.}~\bibnamefont {Li}}, \bibinfo {author} {\bibfnamefont {M.}~\bibnamefont
  {Lu}}, \bibinfo {author} {\bibfnamefont {E.}~\bibnamefont {Megidish}},
  \bibinfo {author} {\bibfnamefont {J.}~\bibnamefont {Marjanovic}}, \bibinfo
  {author} {\bibfnamefont {M.}~\bibnamefont {McDonald}}, \bibinfo {author}
  {\bibfnamefont {T.}~\bibnamefont {Mittiga}}, \bibinfo {author} {\bibfnamefont
  {J.~A.}\ \bibnamefont {Muniz}}, \bibinfo {author} {\bibfnamefont
  {S.}~\bibnamefont {Narayanaswami}}, \bibinfo {author} {\bibfnamefont
  {C.}~\bibnamefont {Nishiguchi}}, \bibinfo {author} {\bibfnamefont
  {R.}~\bibnamefont {Notermans}}, \bibinfo {author} {\bibfnamefont
  {T.}~\bibnamefont {Paule}}, \bibinfo {author} {\bibfnamefont {K.~A.}\
  \bibnamefont {Pawlak}}, \bibinfo {author} {\bibfnamefont {L.~S.}\
  \bibnamefont {Peng}}, \bibinfo {author} {\bibfnamefont {A.}~\bibnamefont
  {Ryou}}, \bibinfo {author} {\bibfnamefont {A.}~\bibnamefont {Smull}},
  \bibinfo {author} {\bibfnamefont {D.}~\bibnamefont {Stack}}, \bibinfo
  {author} {\bibfnamefont {M.}~\bibnamefont {Stone}}, \bibinfo {author}
  {\bibfnamefont {A.}~\bibnamefont {Sucich}}, \bibinfo {author} {\bibfnamefont
  {M.}~\bibnamefont {Urbanek}}, \bibinfo {author} {\bibfnamefont {R.~J.~M.}\
  \bibnamefont {Van De~Veerdonk}}, \bibinfo {author} {\bibfnamefont
  {Z.}~\bibnamefont {Vendeiro}}, \bibinfo {author} {\bibfnamefont
  {T.}~\bibnamefont {Wilkason}}, \bibinfo {author} {\bibfnamefont {T.-Y.}\
  \bibnamefont {Wu}}, \bibinfo {author} {\bibfnamefont {X.}~\bibnamefont
  {Xie}}, \bibinfo {author} {\bibfnamefont {X.}~\bibnamefont {Zhang}},\ and\
  \bibinfo {author} {\bibfnamefont {B.~J.}\ \bibnamefont {Bloom}},\ }\bibfield
  {title} {\bibinfo {title} {Midcircuit {{Qubit Measurement}} and
  {{Rearrangement}} in a {{Yb}} 171 {{Atomic Array}}},\ }\href
  {https://doi.org/10.1103/PhysRevX.13.041034} {\bibfield  {journal} {\bibinfo
  {journal} {Physical Review X}\ }\textbf {\bibinfo {volume} {13}},\ \bibinfo
  {pages} {041034} (\bibinfo {year} {2023})}\BibitemShut {NoStop}%
\bibitem [{\citenamefont {Wu}\ \emph {et~al.}(2022)\citenamefont {Wu},
  \citenamefont {Kolkowitz}, \citenamefont {Puri},\ and\ \citenamefont
  {Thompson}}]{wu2022}%
  \BibitemOpen
  \bibfield  {author} {\bibinfo {author} {\bibfnamefont {Y.}~\bibnamefont
  {Wu}}, \bibinfo {author} {\bibfnamefont {S.}~\bibnamefont {Kolkowitz}},
  \bibinfo {author} {\bibfnamefont {S.}~\bibnamefont {Puri}},\ and\ \bibinfo
  {author} {\bibfnamefont {J.~D.}\ \bibnamefont {Thompson}},\ }\bibfield
  {title} {\bibinfo {title} {Erasure conversion for fault-tolerant quantum
  computing in alkaline earth {{Rydberg}} atom arrays},\ }\href
  {https://doi.org/10.1038/s41467-022-32094-6} {\bibfield  {journal} {\bibinfo
  {journal} {Nature Communications}\ }\textbf {\bibinfo {volume} {13}},\
  \bibinfo {pages} {4657} (\bibinfo {year} {2022})}\BibitemShut {NoStop}%
\bibitem [{\citenamefont {Sahay}\ \emph {et~al.}(2023)\citenamefont {Sahay},
  \citenamefont {Jin}, \citenamefont {Claes}, \citenamefont {Thompson},\ and\
  \citenamefont {Puri}}]{sahay2023a}%
  \BibitemOpen
  \bibfield  {author} {\bibinfo {author} {\bibfnamefont {K.}~\bibnamefont
  {Sahay}}, \bibinfo {author} {\bibfnamefont {J.}~\bibnamefont {Jin}}, \bibinfo
  {author} {\bibfnamefont {J.}~\bibnamefont {Claes}}, \bibinfo {author}
  {\bibfnamefont {J.~D.}\ \bibnamefont {Thompson}},\ and\ \bibinfo {author}
  {\bibfnamefont {S.}~\bibnamefont {Puri}},\ }\bibfield  {title} {\bibinfo
  {title} {High-{{Threshold Codes}} for {{Neutral-Atom Qubits}} with {{Biased
  Erasure Errors}}},\ }\href {https://doi.org/10.1103/PhysRevX.13.041013}
  {\bibfield  {journal} {\bibinfo  {journal} {Physical Review X}\ }\textbf
  {\bibinfo {volume} {13}},\ \bibinfo {pages} {041013} (\bibinfo {year}
  {2023})}\BibitemShut {NoStop}%
\bibitem [{\citenamefont {Huie}\ \emph {et~al.}(2021)\citenamefont {Huie},
  \citenamefont {Menon}, \citenamefont {Bernien},\ and\ \citenamefont
  {Covey}}]{huie2021a}%
  \BibitemOpen
  \bibfield  {author} {\bibinfo {author} {\bibfnamefont {W.}~\bibnamefont
  {Huie}}, \bibinfo {author} {\bibfnamefont {S.~G.}\ \bibnamefont {Menon}},
  \bibinfo {author} {\bibfnamefont {H.}~\bibnamefont {Bernien}},\ and\ \bibinfo
  {author} {\bibfnamefont {J.~P.}\ \bibnamefont {Covey}},\ }\bibfield  {title}
  {\bibinfo {title} {Multiplexed telecommunication-band quantum networking with
  atom arrays in optical cavities},\ }\href
  {https://doi.org/10.1103/PhysRevResearch.3.043154} {\bibfield  {journal}
  {\bibinfo  {journal} {Physical Review Research}\ }\textbf {\bibinfo {volume}
  {3}},\ \bibinfo {pages} {043154} (\bibinfo {year} {2021})}\BibitemShut
  {NoStop}%
\bibitem [{\citenamefont {Young}\ \emph {et~al.}(2022)\citenamefont {Young},
  \citenamefont {Safari}, \citenamefont {Huft}, \citenamefont {Zhang},
  \citenamefont {Oh}, \citenamefont {Chinnarasu},\ and\ \citenamefont
  {Saffman}}]{young2022a}%
  \BibitemOpen
  \bibfield  {author} {\bibinfo {author} {\bibfnamefont {C.~B.}\ \bibnamefont
  {Young}}, \bibinfo {author} {\bibfnamefont {A.}~\bibnamefont {Safari}},
  \bibinfo {author} {\bibfnamefont {P.}~\bibnamefont {Huft}}, \bibinfo {author}
  {\bibfnamefont {J.}~\bibnamefont {Zhang}}, \bibinfo {author} {\bibfnamefont
  {E.}~\bibnamefont {Oh}}, \bibinfo {author} {\bibfnamefont {R.}~\bibnamefont
  {Chinnarasu}},\ and\ \bibinfo {author} {\bibfnamefont {M.}~\bibnamefont
  {Saffman}},\ }\bibfield  {title} {\bibinfo {title} {An architecture for
  quantum networking of neutral atom processors},\ }\href
  {https://doi.org/10.1007/s00340-022-07865-0} {\bibfield  {journal} {\bibinfo
  {journal} {Applied Physics B}\ }\textbf {\bibinfo {volume} {128}},\ \bibinfo
  {pages} {151} (\bibinfo {year} {2022})}\BibitemShut {NoStop}%
\bibitem [{\citenamefont {Saskin}\ \emph {et~al.}(2019)\citenamefont {Saskin},
  \citenamefont {Wilson}, \citenamefont {Grinkemeyer},\ and\ \citenamefont
  {Thompson}}]{Saskin2019}%
  \BibitemOpen
  \bibfield  {author} {\bibinfo {author} {\bibfnamefont {S.}~\bibnamefont
  {Saskin}}, \bibinfo {author} {\bibfnamefont {J.~T.}\ \bibnamefont {Wilson}},
  \bibinfo {author} {\bibfnamefont {B.}~\bibnamefont {Grinkemeyer}},\ and\
  \bibinfo {author} {\bibfnamefont {J.~D.}\ \bibnamefont {Thompson}},\
  }\bibfield  {title} {\bibinfo {title} {Narrow-{{Line Cooling}} and
  {{Imaging}} of {{Ytterbium Atoms}} in an {{Optical Tweezer Array}}},\ }\href
  {https://link.aps.org/doi/10.1103/PhysRevLett.122.143002} {\bibfield
  {journal} {\bibinfo  {journal} {Physical Review Letters}\ }\textbf {\bibinfo
  {volume} {122}},\ \bibinfo {pages} {143002} (\bibinfo {year}
  {2019})}\BibitemShut {NoStop}%
\bibitem [{\citenamefont {Covey}\ \emph {et~al.}(2019)\citenamefont {Covey},
  \citenamefont {Sipahigil}, \citenamefont {Szoke}, \citenamefont {Sinclair},
  \citenamefont {Endres},\ and\ \citenamefont {Painter}}]{covey2019a}%
  \BibitemOpen
  \bibfield  {author} {\bibinfo {author} {\bibfnamefont {J.~P.}\ \bibnamefont
  {Covey}}, \bibinfo {author} {\bibfnamefont {A.}~\bibnamefont {Sipahigil}},
  \bibinfo {author} {\bibfnamefont {S.}~\bibnamefont {Szoke}}, \bibinfo
  {author} {\bibfnamefont {N.}~\bibnamefont {Sinclair}}, \bibinfo {author}
  {\bibfnamefont {M.}~\bibnamefont {Endres}},\ and\ \bibinfo {author}
  {\bibfnamefont {O.}~\bibnamefont {Painter}},\ }\bibfield  {title} {\bibinfo
  {title} {Telecom-{{Band Quantum Optics}} with {{Ytterbium Atoms}} and
  {{Silicon Nanophotonics}}},\ }\href
  {https://doi.org/10.1103/PhysRevApplied.11.034044} {\bibfield  {journal}
  {\bibinfo  {journal} {Physical Review Applied}\ }\textbf {\bibinfo {volume}
  {11}},\ \bibinfo {pages} {034044} (\bibinfo {year} {2019})}\BibitemShut
  {NoStop}%
\bibitem [{\citenamefont {Schine}\ \emph {et~al.}(2016)\citenamefont {Schine},
  \citenamefont {Ryou}, \citenamefont {Gromov}, \citenamefont {Sommer},\ and\
  \citenamefont {Simon}}]{Schine2016nature}%
  \BibitemOpen
  \bibfield  {author} {\bibinfo {author} {\bibfnamefont {N.}~\bibnamefont
  {Schine}}, \bibinfo {author} {\bibfnamefont {A.}~\bibnamefont {Ryou}},
  \bibinfo {author} {\bibfnamefont {A.}~\bibnamefont {Gromov}}, \bibinfo
  {author} {\bibfnamefont {A.}~\bibnamefont {Sommer}},\ and\ \bibinfo {author}
  {\bibfnamefont {J.}~\bibnamefont {Simon}},\ }\bibfield  {title} {\bibinfo
  {title} {Synthetic landau levels for photons},\ }\href
  {https://doi.org/10.1038/nature17943} {\bibfield  {journal} {\bibinfo
  {journal} {Nature}\ }\textbf {\bibinfo {volume} {534}},\ \bibinfo {pages}
  {671} (\bibinfo {year} {2016})}\BibitemShut {NoStop}%
\bibitem [{\citenamefont {Jia}\ \emph {et~al.}(2016)\citenamefont {Jia},
  \citenamefont {Georgakopoulos}, \citenamefont {Ryou}, \citenamefont {Schine},
  \citenamefont {Sommer},\ and\ \citenamefont {Simon}}]{jia2016}%
  \BibitemOpen
  \bibfield  {author} {\bibinfo {author} {\bibfnamefont {N.}~\bibnamefont
  {Jia}}, \bibinfo {author} {\bibfnamefont {A.}~\bibnamefont {Georgakopoulos}},
  \bibinfo {author} {\bibfnamefont {A.}~\bibnamefont {Ryou}}, \bibinfo {author}
  {\bibfnamefont {N.}~\bibnamefont {Schine}}, \bibinfo {author} {\bibfnamefont
  {A.}~\bibnamefont {Sommer}},\ and\ \bibinfo {author} {\bibfnamefont
  {J.}~\bibnamefont {Simon}},\ }\bibfield  {title} {\bibinfo {title}
  {Observation and characterization of cavity {{Rydberg}} polaritons},\ }\href
  {http://link.aps.org/doi/10.1103/PhysRevA.93.041802} {\bibfield  {journal}
  {\bibinfo  {journal} {Physical Review A}\ }\textbf {\bibinfo {volume} {93}},\
  \bibinfo {pages} {41802} (\bibinfo {year} {2016})}\BibitemShut {NoStop}%
\bibitem [{\citenamefont {Cox}\ \emph {et~al.}(2018)\citenamefont {Cox},
  \citenamefont {Meyer}, \citenamefont {Schine}, \citenamefont {Fatemi},\ and\
  \citenamefont {Kunz}}]{cox2018}%
  \BibitemOpen
  \bibfield  {author} {\bibinfo {author} {\bibfnamefont {K.~C.}\ \bibnamefont
  {Cox}}, \bibinfo {author} {\bibfnamefont {D.~H.}\ \bibnamefont {Meyer}},
  \bibinfo {author} {\bibfnamefont {N.~A.}\ \bibnamefont {Schine}}, \bibinfo
  {author} {\bibfnamefont {F.~K.}\ \bibnamefont {Fatemi}},\ and\ \bibinfo
  {author} {\bibfnamefont {P.~D.}\ \bibnamefont {Kunz}},\ }\bibfield  {title}
  {\bibinfo {title} {Increased atom-cavity coupling and stability using a
  parabolic ring cavity},\ }\href {https://doi.org/10.1088/1361-6455/aaddd1}
  {\bibfield  {journal} {\bibinfo  {journal} {Journal of Physics B: Atomic,
  Molecular and Optical Physics}\ }\textbf {\bibinfo {volume} {51}},\ \bibinfo
  {pages} {195002} (\bibinfo {year} {2018})}\BibitemShut {NoStop}%
\bibitem [{\citenamefont {Chen}\ \emph {et~al.}(2022)\citenamefont {Chen},
  \citenamefont {Szurek}, \citenamefont {Hu}, \citenamefont {De~Hond},
  \citenamefont {Braverman},\ and\ \citenamefont {Vuletic}}]{chen2022b}%
  \BibitemOpen
  \bibfield  {author} {\bibinfo {author} {\bibfnamefont {Y.-T.}\ \bibnamefont
  {Chen}}, \bibinfo {author} {\bibfnamefont {M.}~\bibnamefont {Szurek}},
  \bibinfo {author} {\bibfnamefont {B.}~\bibnamefont {Hu}}, \bibinfo {author}
  {\bibfnamefont {J.}~\bibnamefont {De~Hond}}, \bibinfo {author} {\bibfnamefont
  {B.}~\bibnamefont {Braverman}},\ and\ \bibinfo {author} {\bibfnamefont
  {V.}~\bibnamefont {Vuletic}},\ }\bibfield  {title} {\bibinfo {title} {High
  finesse bow-tie cavity for strong atom-photon coupling in {{Rydberg}}
  arrays},\ }\href {https://doi.org/10.1364/OE.469644} {\bibfield  {journal}
  {\bibinfo  {journal} {Optics Express}\ }\textbf {\bibinfo {volume} {30}},\
  \bibinfo {pages} {37426} (\bibinfo {year} {2022})}\BibitemShut {NoStop}%
\bibitem [{\citenamefont {Jia}\ \emph {et~al.}(2018)\citenamefont {Jia},
  \citenamefont {Schine}, \citenamefont {Georgakopoulos}, \citenamefont {Ryou},
  \citenamefont {Sommer},\ and\ \citenamefont {Simon}}]{jia2018}%
  \BibitemOpen
  \bibfield  {author} {\bibinfo {author} {\bibfnamefont {N.}~\bibnamefont
  {Jia}}, \bibinfo {author} {\bibfnamefont {N.}~\bibnamefont {Schine}},
  \bibinfo {author} {\bibfnamefont {A.}~\bibnamefont {Georgakopoulos}},
  \bibinfo {author} {\bibfnamefont {A.}~\bibnamefont {Ryou}}, \bibinfo {author}
  {\bibfnamefont {A.}~\bibnamefont {Sommer}},\ and\ \bibinfo {author}
  {\bibfnamefont {J.}~\bibnamefont {Simon}},\ }\bibfield  {title} {\bibinfo
  {title} {Photons and polaritons in a broken-time-reversal nonplanar
  resonator},\ }\href {https://doi.org/10.1103/PhysRevA.97.013802} {\bibfield
  {journal} {\bibinfo  {journal} {Physical Review A}\ }\textbf {\bibinfo
  {volume} {97}},\ \bibinfo {pages} {013802} (\bibinfo {year}
  {2018})}\BibitemShut {NoStop}%
\bibitem [{\citenamefont {Moehring}\ \emph
  {et~al.}(2007{\natexlab{b}})\citenamefont {Moehring}, \citenamefont {Madsen},
  \citenamefont {Younge}, \citenamefont {Kohn}, \citenamefont {Maunz},
  \citenamefont {Duan}, \citenamefont {Monroe},\ and\ \citenamefont
  {Blinov}}]{moehring2007a}%
  \BibitemOpen
  \bibfield  {author} {\bibinfo {author} {\bibfnamefont {D.~L.}\ \bibnamefont
  {Moehring}}, \bibinfo {author} {\bibfnamefont {M.~J.}\ \bibnamefont
  {Madsen}}, \bibinfo {author} {\bibfnamefont {K.~C.}\ \bibnamefont {Younge}},
  \bibinfo {author} {\bibfnamefont {R.~N.}\ \bibnamefont {Kohn}, \bibfnamefont
  {Jr.}}, \bibinfo {author} {\bibfnamefont {P.}~\bibnamefont {Maunz}}, \bibinfo
  {author} {\bibfnamefont {L.-M.}\ \bibnamefont {Duan}}, \bibinfo {author}
  {\bibfnamefont {C.}~\bibnamefont {Monroe}},\ and\ \bibinfo {author}
  {\bibfnamefont {B.~B.}\ \bibnamefont {Blinov}},\ }\bibfield  {title}
  {\bibinfo {title} {Quantum networking with photons and trapped atoms
  ({{Invited}})},\ }\href {https://doi.org/10.1364/JOSAB.24.000300} {\bibfield
  {journal} {\bibinfo  {journal} {Journal of the Optical Society of America B}\
  }\textbf {\bibinfo {volume} {24}},\ \bibinfo {pages} {300} (\bibinfo {year}
  {2007}{\natexlab{b}})}\BibitemShut {NoStop}%
\bibitem [{\citenamefont {Bluvstein}\ \emph {et~al.}(2022)\citenamefont
  {Bluvstein}, \citenamefont {Levine}, \citenamefont {Semeghini}, \citenamefont
  {Wang}, \citenamefont {Ebadi}, \citenamefont {Kalinowski}, \citenamefont
  {Keesling}, \citenamefont {Maskara}, \citenamefont {Pichler}, \citenamefont
  {Greiner}, \citenamefont {Vuleti{\'c}},\ and\ \citenamefont
  {Lukin}}]{bluvstein2022}%
  \BibitemOpen
  \bibfield  {author} {\bibinfo {author} {\bibfnamefont {D.}~\bibnamefont
  {Bluvstein}}, \bibinfo {author} {\bibfnamefont {H.}~\bibnamefont {Levine}},
  \bibinfo {author} {\bibfnamefont {G.}~\bibnamefont {Semeghini}}, \bibinfo
  {author} {\bibfnamefont {T.~T.}\ \bibnamefont {Wang}}, \bibinfo {author}
  {\bibfnamefont {S.}~\bibnamefont {Ebadi}}, \bibinfo {author} {\bibfnamefont
  {M.}~\bibnamefont {Kalinowski}}, \bibinfo {author} {\bibfnamefont
  {A.}~\bibnamefont {Keesling}}, \bibinfo {author} {\bibfnamefont
  {N.}~\bibnamefont {Maskara}}, \bibinfo {author} {\bibfnamefont
  {H.}~\bibnamefont {Pichler}}, \bibinfo {author} {\bibfnamefont
  {M.}~\bibnamefont {Greiner}}, \bibinfo {author} {\bibfnamefont
  {V.}~\bibnamefont {Vuleti{\'c}}},\ and\ \bibinfo {author} {\bibfnamefont
  {M.~D.}\ \bibnamefont {Lukin}},\ }\bibfield  {title} {\bibinfo {title} {A
  quantum processor based on coherent transport of entangled atom arrays},\
  }\href {https://doi.org/10.1038/s41586-022-04592-6} {\bibfield  {journal}
  {\bibinfo  {journal} {Nature}\ }\textbf {\bibinfo {volume} {604}},\ \bibinfo
  {pages} {451} (\bibinfo {year} {2022})}\BibitemShut {NoStop}%
\bibitem [{\citenamefont {Vasilev}\ \emph {et~al.}(2010)\citenamefont
  {Vasilev}, \citenamefont {Ljunggren},\ and\ \citenamefont
  {Kuhn}}]{vasilev2010}%
  \BibitemOpen
  \bibfield  {author} {\bibinfo {author} {\bibfnamefont {G.~S.}\ \bibnamefont
  {Vasilev}}, \bibinfo {author} {\bibfnamefont {D.}~\bibnamefont {Ljunggren}},\
  and\ \bibinfo {author} {\bibfnamefont {A.}~\bibnamefont {Kuhn}},\ }\bibfield
  {title} {\bibinfo {title} {Single photons made-to-measure},\ }\href
  {https://doi.org/10.1088/1367-2630/12/6/063024} {\bibfield  {journal}
  {\bibinfo  {journal} {New Journal of Physics}\ }\textbf {\bibinfo {volume}
  {12}},\ \bibinfo {pages} {063024} (\bibinfo {year} {2010})}\BibitemShut
  {NoStop}%
\bibitem [{\citenamefont {Timurdogan}\ \emph {et~al.}(2019)\citenamefont
  {Timurdogan}, \citenamefont {Su}, \citenamefont {Shiue}, \citenamefont
  {Poulton}, \citenamefont {Byrd}, \citenamefont {Xin},\ and\ \citenamefont
  {Watts}}]{timurdogan2019}%
  \BibitemOpen
  \bibfield  {author} {\bibinfo {author} {\bibfnamefont {E.}~\bibnamefont
  {Timurdogan}}, \bibinfo {author} {\bibfnamefont {Z.}~\bibnamefont {Su}},
  \bibinfo {author} {\bibfnamefont {R.-J.}\ \bibnamefont {Shiue}}, \bibinfo
  {author} {\bibfnamefont {C.~V.}\ \bibnamefont {Poulton}}, \bibinfo {author}
  {\bibfnamefont {M.~J.}\ \bibnamefont {Byrd}}, \bibinfo {author}
  {\bibfnamefont {S.}~\bibnamefont {Xin}},\ and\ \bibinfo {author}
  {\bibfnamefont {M.~R.}\ \bibnamefont {Watts}},\ }\bibfield  {title} {\bibinfo
  {title} {{{APSUNY Process Design Kit}} ({{PDKv3}}.0): {{O}}, {{C}} and {{L
  Band Silicon Photonics Component Libraries}} on 300mm {{Wafers}}},\ }in\
  \href {https://ieeexplore.ieee.org/document/8696994} {\emph {\bibinfo
  {booktitle} {2019 {{Optical Fiber Communications Conference}} and
  {{Exhibition}} ({{OFC}})}}}\ (\bibinfo {year} {2019})\ pp.\ \bibinfo {pages}
  {1--3}\BibitemShut {NoStop}%
\bibitem [{\citenamefont {Ma}\ \emph {et~al.}(2022)\citenamefont {Ma},
  \citenamefont {Burgers}, \citenamefont {Liu}, \citenamefont {Wilson},
  \citenamefont {Zhang},\ and\ \citenamefont {Thompson}}]{ma2022}%
  \BibitemOpen
  \bibfield  {author} {\bibinfo {author} {\bibfnamefont {S.}~\bibnamefont
  {Ma}}, \bibinfo {author} {\bibfnamefont {A.~P.}\ \bibnamefont {Burgers}},
  \bibinfo {author} {\bibfnamefont {G.}~\bibnamefont {Liu}}, \bibinfo {author}
  {\bibfnamefont {J.}~\bibnamefont {Wilson}}, \bibinfo {author} {\bibfnamefont
  {B.}~\bibnamefont {Zhang}},\ and\ \bibinfo {author} {\bibfnamefont {J.~D.}\
  \bibnamefont {Thompson}},\ }\bibfield  {title} {\bibinfo {title} {Universal
  {{Gate Operations}} on {{Nuclear Spin Qubits}} in an {{Optical Tweezer
  Array}} of {{Yb}} 171 {{Atoms}}},\ }\href
  {https://doi.org/10.1103/PhysRevX.12.021028} {\bibfield  {journal} {\bibinfo
  {journal} {Physical Review X}\ }\textbf {\bibinfo {volume} {12}},\ \bibinfo
  {pages} {021028} (\bibinfo {year} {2022})}\BibitemShut {NoStop}%
\bibitem [{\citenamefont {Jenkins}\ \emph {et~al.}(2022)\citenamefont
  {Jenkins}, \citenamefont {Lis}, \citenamefont {Senoo}, \citenamefont
  {McGrew},\ and\ \citenamefont {Kaufman}}]{jenkins2022}%
  \BibitemOpen
  \bibfield  {author} {\bibinfo {author} {\bibfnamefont {A.}~\bibnamefont
  {Jenkins}}, \bibinfo {author} {\bibfnamefont {J.~W.}\ \bibnamefont {Lis}},
  \bibinfo {author} {\bibfnamefont {A.}~\bibnamefont {Senoo}}, \bibinfo
  {author} {\bibfnamefont {W.~F.}\ \bibnamefont {McGrew}},\ and\ \bibinfo
  {author} {\bibfnamefont {A.~M.}\ \bibnamefont {Kaufman}},\ }\bibfield
  {title} {\bibinfo {title} {Ytterbium {{Nuclear-Spin Qubits}} in an {{Optical
  Tweezer Array}}},\ }\href {https://doi.org/10.1103/PhysRevX.12.021027}
  {\bibfield  {journal} {\bibinfo  {journal} {Physical Review X}\ }\textbf
  {\bibinfo {volume} {12}},\ \bibinfo {pages} {021027} (\bibinfo {year}
  {2022})}\BibitemShut {NoStop}%
\bibitem [{\citenamefont {Merkel}\ \emph {et~al.}(2019)\citenamefont {Merkel},
  \citenamefont {Thirumalai}, \citenamefont {Tarlton}, \citenamefont
  {Sch{\"a}fer}, \citenamefont {Ballance}, \citenamefont {Harty},\ and\
  \citenamefont {Lucas}}]{merkel2019}%
  \BibitemOpen
  \bibfield  {author} {\bibinfo {author} {\bibfnamefont {B.}~\bibnamefont
  {Merkel}}, \bibinfo {author} {\bibfnamefont {K.}~\bibnamefont {Thirumalai}},
  \bibinfo {author} {\bibfnamefont {J.~E.}\ \bibnamefont {Tarlton}}, \bibinfo
  {author} {\bibfnamefont {V.~M.}\ \bibnamefont {Sch{\"a}fer}}, \bibinfo
  {author} {\bibfnamefont {C.~J.}\ \bibnamefont {Ballance}}, \bibinfo {author}
  {\bibfnamefont {T.~P.}\ \bibnamefont {Harty}},\ and\ \bibinfo {author}
  {\bibfnamefont {D.~M.}\ \bibnamefont {Lucas}},\ }\bibfield  {title} {\bibinfo
  {title} {Magnetic field stabilization system for atomic physics
  experiments},\ }\href {https://doi.org/10.1063/1.5080093} {\bibfield
  {journal} {\bibinfo  {journal} {Review of Scientific Instruments}\ }\textbf
  {\bibinfo {volume} {90}},\ \bibinfo {pages} {044702} (\bibinfo {year}
  {2019})}\BibitemShut {NoStop}%
\bibitem [{\citenamefont {Tiecke}\ \emph {et~al.}(2014)\citenamefont {Tiecke},
  \citenamefont {Thompson}, \citenamefont {{de Leon}}, \citenamefont {Liu},
  \citenamefont {Vuleti{\'c}},\ and\ \citenamefont {Lukin}}]{tiecke2014}%
  \BibitemOpen
  \bibfield  {author} {\bibinfo {author} {\bibfnamefont {T.~G.}\ \bibnamefont
  {Tiecke}}, \bibinfo {author} {\bibfnamefont {J.~D.}\ \bibnamefont
  {Thompson}}, \bibinfo {author} {\bibfnamefont {N.~P.}\ \bibnamefont {{de
  Leon}}}, \bibinfo {author} {\bibfnamefont {L.~R.}\ \bibnamefont {Liu}},
  \bibinfo {author} {\bibfnamefont {V.}~\bibnamefont {Vuleti{\'c}}},\ and\
  \bibinfo {author} {\bibfnamefont {M.~D.}\ \bibnamefont {Lukin}},\ }\bibfield
  {title} {\bibinfo {title} {Nanophotonic quantum phase switch with a single
  atom},\ }\href {http://www.nature.com/doifinder/10.1038/nature13188}
  {\bibfield  {journal} {\bibinfo  {journal} {Nature}\ }\textbf {\bibinfo
  {volume} {508}},\ \bibinfo {pages} {241} (\bibinfo {year}
  {2014})}\BibitemShut {NoStop}%
\bibitem [{\citenamefont {Schine}\ \emph {et~al.}(2022)\citenamefont {Schine},
  \citenamefont {Young}, \citenamefont {Eckner}, \citenamefont {Martin},\ and\
  \citenamefont {Kaufman}}]{schine2022}%
  \BibitemOpen
  \bibfield  {author} {\bibinfo {author} {\bibfnamefont {N.}~\bibnamefont
  {Schine}}, \bibinfo {author} {\bibfnamefont {A.~W.}\ \bibnamefont {Young}},
  \bibinfo {author} {\bibfnamefont {W.~J.}\ \bibnamefont {Eckner}}, \bibinfo
  {author} {\bibfnamefont {M.~J.}\ \bibnamefont {Martin}},\ and\ \bibinfo
  {author} {\bibfnamefont {A.~M.}\ \bibnamefont {Kaufman}},\ }\bibfield
  {title} {\bibinfo {title} {Long-lived {{Bell}} states in an array of optical
  clock qubits},\ }\href {https://doi.org/10.1038/s41567-022-01678-w}
  {\bibfield  {journal} {\bibinfo  {journal} {Nature Physics}\ }\textbf
  {\bibinfo {volume} {18}},\ \bibinfo {pages} {1067} (\bibinfo {year}
  {2022})}\BibitemShut {NoStop}%
\bibitem [{\citenamefont {Shibata}\ \emph {et~al.}(2015)\citenamefont
  {Shibata}, \citenamefont {Shimizu}, \citenamefont {Takesue},\ and\
  \citenamefont {Tokura}}]{shibata2015}%
  \BibitemOpen
  \bibfield  {author} {\bibinfo {author} {\bibfnamefont {H.}~\bibnamefont
  {Shibata}}, \bibinfo {author} {\bibfnamefont {K.}~\bibnamefont {Shimizu}},
  \bibinfo {author} {\bibfnamefont {H.}~\bibnamefont {Takesue}},\ and\ \bibinfo
  {author} {\bibfnamefont {Y.}~\bibnamefont {Tokura}},\ }\bibfield  {title}
  {\bibinfo {title} {Ultimate low system dark-count rate for superconducting
  nanowire single-photon detector},\ }\href
  {https://doi.org/10.1364/OL.40.003428} {\bibfield  {journal} {\bibinfo
  {journal} {Optics Letters}\ }\textbf {\bibinfo {volume} {40}},\ \bibinfo
  {pages} {3428} (\bibinfo {year} {2015})}\BibitemShut {NoStop}%
\bibitem [{\citenamefont {Schuck}\ \emph {et~al.}(2013)\citenamefont {Schuck},
  \citenamefont {Pernice},\ and\ \citenamefont {Tang}}]{schuck2013}%
  \BibitemOpen
  \bibfield  {author} {\bibinfo {author} {\bibfnamefont {C.}~\bibnamefont
  {Schuck}}, \bibinfo {author} {\bibfnamefont {W.~H.~P.}\ \bibnamefont
  {Pernice}},\ and\ \bibinfo {author} {\bibfnamefont {H.~X.}\ \bibnamefont
  {Tang}},\ }\bibfield  {title} {\bibinfo {title} {Waveguide integrated low
  noise {{NbTiN}} nanowire single-photon detectors with milli-{{Hz}} dark count
  rate},\ }\href
  {papers3://publication/uuid/1D8F159C-9A5A-4886-8833-4A1043EFB7AB} {\bibfield
  {journal} {\bibinfo  {journal} {Scientific Reports}\ }\textbf {\bibinfo
  {volume} {3}},\ \bibinfo {pages} {1893} (\bibinfo {year} {2013})}\BibitemShut
  {NoStop}%
\bibitem [{\citenamefont {Burgers}\ \emph {et~al.}(2022)\citenamefont
  {Burgers}, \citenamefont {Ma}, \citenamefont {Saskin}, \citenamefont
  {Wilson}, \citenamefont {Alarc{\'o}n}, \citenamefont {Greene},\ and\
  \citenamefont {Thompson}}]{burgers2022}%
  \BibitemOpen
  \bibfield  {author} {\bibinfo {author} {\bibfnamefont {A.~P.}\ \bibnamefont
  {Burgers}}, \bibinfo {author} {\bibfnamefont {S.}~\bibnamefont {Ma}},
  \bibinfo {author} {\bibfnamefont {S.}~\bibnamefont {Saskin}}, \bibinfo
  {author} {\bibfnamefont {J.}~\bibnamefont {Wilson}}, \bibinfo {author}
  {\bibfnamefont {M.~A.}\ \bibnamefont {Alarc{\'o}n}}, \bibinfo {author}
  {\bibfnamefont {C.~H.}\ \bibnamefont {Greene}},\ and\ \bibinfo {author}
  {\bibfnamefont {J.~D.}\ \bibnamefont {Thompson}},\ }\bibfield  {title}
  {\bibinfo {title} {Controlling {{Rydberg Excitations Using Ion-Core
  Transitions}} in {{Alkaline-Earth Atom-Tweezer Arrays}}},\ }\href
  {https://doi.org/10.1103/PRXQuantum.3.020326} {\bibfield  {journal} {\bibinfo
   {journal} {PRX Quantum}\ }\textbf {\bibinfo {volume} {3}},\ \bibinfo {pages}
  {020326} (\bibinfo {year} {2022})}\BibitemShut {NoStop}%
\bibitem [{\citenamefont {Zhang}\ \emph {et~al.}(2023)\citenamefont {Zhang},
  \citenamefont {Peng}, \citenamefont {Paul},\ and\ \citenamefont
  {Thompson}}]{zhang2023a}%
  \BibitemOpen
  \bibfield  {author} {\bibinfo {author} {\bibfnamefont {B.}~\bibnamefont
  {Zhang}}, \bibinfo {author} {\bibfnamefont {P.}~\bibnamefont {Peng}},
  \bibinfo {author} {\bibfnamefont {A.}~\bibnamefont {Paul}},\ and\ \bibinfo
  {author} {\bibfnamefont {J.~D.}\ \bibnamefont {Thompson}},\ }\href
  {http://arxiv.org/abs/2310.08539} {\bibinfo {title} {A scaled local gate
  controller for optically addressed qubits}} (\bibinfo {year} {2023}),\
  \Eprint {https://arxiv.org/abs/2310.08539} {arxiv:2310.08539 [quant-ph]}
  \BibitemShut {NoStop}%
\bibitem [{\citenamefont {Aliferis}\ and\ \citenamefont
  {Preskill}(2008)}]{Aliferis2008}%
  \BibitemOpen
  \bibfield  {author} {\bibinfo {author} {\bibfnamefont {P.}~\bibnamefont
  {Aliferis}}\ and\ \bibinfo {author} {\bibfnamefont {J.}~\bibnamefont
  {Preskill}},\ }\bibfield  {title} {\bibinfo {title} {Fault-tolerant quantum
  computation against biased noise},\ }\href
  {https://doi.org/10.1103/PhysRevA.78.052331} {\bibfield  {journal} {\bibinfo
  {journal} {Physical Review A}\ }\textbf {\bibinfo {volume} {78}},\ \bibinfo
  {pages} {052331} (\bibinfo {year} {2008})}\BibitemShut {NoStop}%
\bibitem [{\citenamefont {Bonilla~Ataides}\ \emph {et~al.}(2021)\citenamefont
  {Bonilla~Ataides}, \citenamefont {Tuckett}, \citenamefont {Bartlett},
  \citenamefont {Flammia},\ and\ \citenamefont {Brown}}]{BonillaAtaides2021}%
  \BibitemOpen
  \bibfield  {author} {\bibinfo {author} {\bibfnamefont {J.~P.}\ \bibnamefont
  {Bonilla~Ataides}}, \bibinfo {author} {\bibfnamefont {D.~K.}\ \bibnamefont
  {Tuckett}}, \bibinfo {author} {\bibfnamefont {S.~D.}\ \bibnamefont
  {Bartlett}}, \bibinfo {author} {\bibfnamefont {S.~T.}\ \bibnamefont
  {Flammia}},\ and\ \bibinfo {author} {\bibfnamefont {B.~J.}\ \bibnamefont
  {Brown}},\ }\bibfield  {title} {\bibinfo {title} {The xzzx surface code},\
  }\href {https://doi.org/10.1038/s41467-021-22274-1} {\bibfield  {journal}
  {\bibinfo  {journal} {Nature Communications}\ }\textbf {\bibinfo {volume}
  {12}},\ \bibinfo {pages} {2172} (\bibinfo {year} {2021})}\BibitemShut
  {NoStop}%
\bibitem [{\citenamefont {Grassl}\ \emph {et~al.}(1997)\citenamefont {Grassl},
  \citenamefont {Beth},\ and\ \citenamefont {Pellizzari}}]{grassl1997}%
  \BibitemOpen
  \bibfield  {author} {\bibinfo {author} {\bibfnamefont {M.}~\bibnamefont
  {Grassl}}, \bibinfo {author} {\bibfnamefont {{\relax Th}.}~\bibnamefont
  {Beth}},\ and\ \bibinfo {author} {\bibfnamefont {T.}~\bibnamefont
  {Pellizzari}},\ }\bibfield  {title} {\bibinfo {title} {Codes for the quantum
  erasure channel},\ }\href {https://doi.org/10.1103/PhysRevA.56.33} {\bibfield
   {journal} {\bibinfo  {journal} {Physical Review A}\ }\textbf {\bibinfo
  {volume} {56}},\ \bibinfo {pages} {33} (\bibinfo {year} {1997})}\BibitemShut
  {NoStop}%
\bibitem [{\citenamefont {Stace}\ \emph {et~al.}(2009)\citenamefont {Stace},
  \citenamefont {Barrett},\ and\ \citenamefont {Doherty}}]{stace2009}%
  \BibitemOpen
  \bibfield  {author} {\bibinfo {author} {\bibfnamefont {T.~M.}\ \bibnamefont
  {Stace}}, \bibinfo {author} {\bibfnamefont {S.~D.}\ \bibnamefont {Barrett}},\
  and\ \bibinfo {author} {\bibfnamefont {A.~C.}\ \bibnamefont {Doherty}},\
  }\bibfield  {title} {\bibinfo {title} {Thresholds for {{Topological Codes}}
  in the {{Presence}} of {{Loss}}},\ }\href
  {https://doi.org/10.1103/PhysRevLett.102.200501} {\bibfield  {journal}
  {\bibinfo  {journal} {Physical Review Letters}\ }\textbf {\bibinfo {volume}
  {102}},\ \bibinfo {pages} {200501} (\bibinfo {year} {2009})}\BibitemShut
  {NoStop}%
\bibitem [{\citenamefont {Pattison}\ \emph {et~al.}(2021)\citenamefont
  {Pattison}, \citenamefont {Beverland}, \citenamefont {{da Silva}},\ and\
  \citenamefont {Delfosse}}]{pattison2021}%
  \BibitemOpen
  \bibfield  {author} {\bibinfo {author} {\bibfnamefont {C.~A.}\ \bibnamefont
  {Pattison}}, \bibinfo {author} {\bibfnamefont {M.~E.}\ \bibnamefont
  {Beverland}}, \bibinfo {author} {\bibfnamefont {M.~P.}\ \bibnamefont {{da
  Silva}}},\ and\ \bibinfo {author} {\bibfnamefont {N.}~\bibnamefont
  {Delfosse}},\ }\href {http://arxiv.org/abs/2107.13589} {\bibinfo {title}
  {Improved quantum error correction using soft information}} (\bibinfo {year}
  {2021}),\ \Eprint {https://arxiv.org/abs/2107.13589} {arxiv:2107.13589
  [quant-ph]} \BibitemShut {NoStop}%
\bibitem [{\citenamefont {Zhao}\ \emph {et~al.}(2014)\citenamefont {Zhao},
  \citenamefont {Zhang}, \citenamefont {Yang}, \citenamefont {Sang},
  \citenamefont {Jiang}, \citenamefont {Bao},\ and\ \citenamefont
  {Pan}}]{zhao2014}%
  \BibitemOpen
  \bibfield  {author} {\bibinfo {author} {\bibfnamefont {T.-M.}\ \bibnamefont
  {Zhao}}, \bibinfo {author} {\bibfnamefont {H.}~\bibnamefont {Zhang}},
  \bibinfo {author} {\bibfnamefont {J.}~\bibnamefont {Yang}}, \bibinfo {author}
  {\bibfnamefont {Z.-R.}\ \bibnamefont {Sang}}, \bibinfo {author}
  {\bibfnamefont {X.}~\bibnamefont {Jiang}}, \bibinfo {author} {\bibfnamefont
  {X.-H.}\ \bibnamefont {Bao}},\ and\ \bibinfo {author} {\bibfnamefont {J.-W.}\
  \bibnamefont {Pan}},\ }\bibfield  {title} {\bibinfo {title} {Entangling
  different-color photons via time-resolved measurement and active feed
  forward},\ }\href {https://doi.org/10.1103/PhysRevLett.112.103602} {\bibfield
   {journal} {\bibinfo  {journal} {Physical Review Letters}\ }\textbf {\bibinfo
  {volume} {112}},\ \bibinfo {pages} {103602} (\bibinfo {year}
  {2014})}\BibitemShut {NoStop}%
\bibitem [{\citenamefont {Menssen}\ \emph {et~al.}(2023)\citenamefont
  {Menssen}, \citenamefont {Hermans}, \citenamefont {Christen}, \citenamefont
  {Propson}, \citenamefont {Li}, \citenamefont {Leenheer}, \citenamefont
  {Zimmermann}, \citenamefont {Dong}, \citenamefont {Larocque}, \citenamefont
  {Raniwala}, \citenamefont {Gilbert}, \citenamefont {Eichenfield},\ and\
  \citenamefont {Englund}}]{menssen2023}%
  \BibitemOpen
  \bibfield  {author} {\bibinfo {author} {\bibfnamefont {A.~J.}\ \bibnamefont
  {Menssen}}, \bibinfo {author} {\bibfnamefont {A.}~\bibnamefont {Hermans}},
  \bibinfo {author} {\bibfnamefont {I.}~\bibnamefont {Christen}}, \bibinfo
  {author} {\bibfnamefont {T.}~\bibnamefont {Propson}}, \bibinfo {author}
  {\bibfnamefont {C.}~\bibnamefont {Li}}, \bibinfo {author} {\bibfnamefont
  {A.~J.}\ \bibnamefont {Leenheer}}, \bibinfo {author} {\bibfnamefont
  {M.}~\bibnamefont {Zimmermann}}, \bibinfo {author} {\bibfnamefont
  {M.}~\bibnamefont {Dong}}, \bibinfo {author} {\bibfnamefont {H.}~\bibnamefont
  {Larocque}}, \bibinfo {author} {\bibfnamefont {H.}~\bibnamefont {Raniwala}},
  \bibinfo {author} {\bibfnamefont {G.}~\bibnamefont {Gilbert}}, \bibinfo
  {author} {\bibfnamefont {M.}~\bibnamefont {Eichenfield}},\ and\ \bibinfo
  {author} {\bibfnamefont {D.~R.}\ \bibnamefont {Englund}},\ }\bibfield
  {title} {\bibinfo {title} {Scalable photonic integrated circuits for
  high-fidelity light control},\ }\href {https://doi.org/10.1364/OPTICA.489504}
  {\bibfield  {journal} {\bibinfo  {journal} {Optica}\ }\textbf {\bibinfo
  {volume} {10}},\ \bibinfo {pages} {1366} (\bibinfo {year}
  {2023})}\BibitemShut {NoStop}%
\bibitem [{\citenamefont {Jiang}\ \emph {et~al.}(2007)\citenamefont {Jiang},
  \citenamefont {Taylor}, \citenamefont {S{\o}rensen},\ and\ \citenamefont
  {Lukin}}]{jiang2007}%
  \BibitemOpen
  \bibfield  {author} {\bibinfo {author} {\bibfnamefont {L.}~\bibnamefont
  {Jiang}}, \bibinfo {author} {\bibfnamefont {J.~M.}\ \bibnamefont {Taylor}},
  \bibinfo {author} {\bibfnamefont {A.~S.}\ \bibnamefont {S{\o}rensen}},\ and\
  \bibinfo {author} {\bibfnamefont {M.~D.}\ \bibnamefont {Lukin}},\ }\bibfield
  {title} {\bibinfo {title} {Distributed quantum computation based on small
  quantum registers},\ }\href {https://doi.org/10.1103/PhysRevA.76.062323}
  {\bibfield  {journal} {\bibinfo  {journal} {Physical Review A}\ }\textbf
  {\bibinfo {volume} {76}},\ \bibinfo {pages} {062323} (\bibinfo {year}
  {2007})}\BibitemShut {NoStop}%
\bibitem [{\citenamefont {Ebadi}\ \emph {et~al.}(2021)\citenamefont {Ebadi},
  \citenamefont {Wang}, \citenamefont {Levine}, \citenamefont {Keesling},
  \citenamefont {Semeghini}, \citenamefont {Omran}, \citenamefont {Bluvstein},
  \citenamefont {Samajdar}, \citenamefont {Pichler}, \citenamefont {Ho},
  \citenamefont {Choi}, \citenamefont {Sachdev}, \citenamefont {Greiner},
  \citenamefont {Vuletić},\ and\ \citenamefont {Lukin}}]{ebadi2021}%
  \BibitemOpen
  \bibfield  {author} {\bibinfo {author} {\bibfnamefont {S.}~\bibnamefont
  {Ebadi}}, \bibinfo {author} {\bibfnamefont {T.~T.}\ \bibnamefont {Wang}},
  \bibinfo {author} {\bibfnamefont {H.}~\bibnamefont {Levine}}, \bibinfo
  {author} {\bibfnamefont {A.}~\bibnamefont {Keesling}}, \bibinfo {author}
  {\bibfnamefont {G.}~\bibnamefont {Semeghini}}, \bibinfo {author}
  {\bibfnamefont {A.}~\bibnamefont {Omran}}, \bibinfo {author} {\bibfnamefont
  {D.}~\bibnamefont {Bluvstein}}, \bibinfo {author} {\bibfnamefont
  {R.}~\bibnamefont {Samajdar}}, \bibinfo {author} {\bibfnamefont
  {H.}~\bibnamefont {Pichler}}, \bibinfo {author} {\bibfnamefont {W.~W.}\
  \bibnamefont {Ho}}, \bibinfo {author} {\bibfnamefont {S.}~\bibnamefont
  {Choi}}, \bibinfo {author} {\bibfnamefont {S.}~\bibnamefont {Sachdev}},
  \bibinfo {author} {\bibfnamefont {M.}~\bibnamefont {Greiner}}, \bibinfo
  {author} {\bibfnamefont {V.}~\bibnamefont {Vuletić}},\ and\ \bibinfo
  {author} {\bibfnamefont {M.~D.}\ \bibnamefont {Lukin}},\ }\bibfield  {title}
  {\bibinfo {title} {Quantum phases of matter on a 256-atom programmable
  quantum simulator},\ }\href {https://doi.org/10.1038/s41586-021-03582-4}
  {\bibfield  {journal} {\bibinfo  {journal} {Nature}\ }\textbf {\bibinfo
  {volume} {595}},\ \bibinfo {pages} {227} (\bibinfo {year}
  {2021})}\BibitemShut {NoStop}%
\bibitem [{\citenamefont {Scholl}\ \emph {et~al.}(2021)\citenamefont {Scholl},
  \citenamefont {Schuler}, \citenamefont {Williams}, \citenamefont
  {Eberharter}, \citenamefont {Barredo}, \citenamefont {Schymik}, \citenamefont
  {Lienhard}, \citenamefont {Henry}, \citenamefont {Lang}, \citenamefont
  {Lahaye}, \citenamefont {Läuchli},\ and\ \citenamefont
  {Browaeys}}]{scholl2021}%
  \BibitemOpen
  \bibfield  {author} {\bibinfo {author} {\bibfnamefont {P.}~\bibnamefont
  {Scholl}}, \bibinfo {author} {\bibfnamefont {M.}~\bibnamefont {Schuler}},
  \bibinfo {author} {\bibfnamefont {H.~J.}\ \bibnamefont {Williams}}, \bibinfo
  {author} {\bibfnamefont {A.~A.}\ \bibnamefont {Eberharter}}, \bibinfo
  {author} {\bibfnamefont {D.}~\bibnamefont {Barredo}}, \bibinfo {author}
  {\bibfnamefont {K.-N.}\ \bibnamefont {Schymik}}, \bibinfo {author}
  {\bibfnamefont {V.}~\bibnamefont {Lienhard}}, \bibinfo {author}
  {\bibfnamefont {L.-P.}\ \bibnamefont {Henry}}, \bibinfo {author}
  {\bibfnamefont {T.~C.}\ \bibnamefont {Lang}}, \bibinfo {author}
  {\bibfnamefont {T.}~\bibnamefont {Lahaye}}, \bibinfo {author} {\bibfnamefont
  {A.~M.}\ \bibnamefont {Läuchli}},\ and\ \bibinfo {author} {\bibfnamefont
  {A.}~\bibnamefont {Browaeys}},\ }\bibfield  {title} {\bibinfo {title}
  {Quantum simulation of 2d antiferromagnets with hundreds of rydberg atoms},\
  }\href {https://doi.org/10.1038/s41586-021-03585-1} {\bibfield  {journal}
  {\bibinfo  {journal} {Nature}\ }\textbf {\bibinfo {volume} {595}},\ \bibinfo
  {pages} {233} (\bibinfo {year} {2021})}\BibitemShut {NoStop}%
\bibitem [{\citenamefont {Scholl}\ \emph {et~al.}(2023)\citenamefont {Scholl},
  \citenamefont {Shaw}, \citenamefont {Tsai}, \citenamefont {Finkelstein},
  \citenamefont {Choi},\ and\ \citenamefont {Endres}}]{scholl2023a}%
  \BibitemOpen
  \bibfield  {author} {\bibinfo {author} {\bibfnamefont {P.}~\bibnamefont
  {Scholl}}, \bibinfo {author} {\bibfnamefont {A.~L.}\ \bibnamefont {Shaw}},
  \bibinfo {author} {\bibfnamefont {R.~B.-S.}\ \bibnamefont {Tsai}}, \bibinfo
  {author} {\bibfnamefont {R.}~\bibnamefont {Finkelstein}}, \bibinfo {author}
  {\bibfnamefont {J.}~\bibnamefont {Choi}},\ and\ \bibinfo {author}
  {\bibfnamefont {M.}~\bibnamefont {Endres}},\ }\bibfield  {title} {\bibinfo
  {title} {Erasure conversion in a high-fidelity {{Rydberg}} quantum
  simulator},\ }\href {https://doi.org/10.1038/s41586-023-06516-4} {\bibfield
  {journal} {\bibinfo  {journal} {Nature}\ }\textbf {\bibinfo {volume} {622}},\
  \bibinfo {pages} {273} (\bibinfo {year} {2023})}\BibitemShut {NoStop}%
\bibitem [{\citenamefont {Evered}\ \emph {et~al.}(2023)\citenamefont {Evered},
  \citenamefont {Bluvstein}, \citenamefont {Kalinowski}, \citenamefont {Ebadi},
  \citenamefont {Manovitz}, \citenamefont {Zhou}, \citenamefont {Li},
  \citenamefont {Geim}, \citenamefont {Wang}, \citenamefont {Maskara},
  \citenamefont {Levine}, \citenamefont {Semeghini}, \citenamefont {Greiner},
  \citenamefont {Vuletić},\ and\ \citenamefont {Lukin}}]{evered2023}%
  \BibitemOpen
  \bibfield  {author} {\bibinfo {author} {\bibfnamefont {S.~J.}\ \bibnamefont
  {Evered}}, \bibinfo {author} {\bibfnamefont {D.}~\bibnamefont {Bluvstein}},
  \bibinfo {author} {\bibfnamefont {M.}~\bibnamefont {Kalinowski}}, \bibinfo
  {author} {\bibfnamefont {S.}~\bibnamefont {Ebadi}}, \bibinfo {author}
  {\bibfnamefont {T.}~\bibnamefont {Manovitz}}, \bibinfo {author}
  {\bibfnamefont {H.}~\bibnamefont {Zhou}}, \bibinfo {author} {\bibfnamefont
  {S.~H.}\ \bibnamefont {Li}}, \bibinfo {author} {\bibfnamefont {A.~A.}\
  \bibnamefont {Geim}}, \bibinfo {author} {\bibfnamefont {T.~T.}\ \bibnamefont
  {Wang}}, \bibinfo {author} {\bibfnamefont {N.}~\bibnamefont {Maskara}},
  \bibinfo {author} {\bibfnamefont {H.}~\bibnamefont {Levine}}, \bibinfo
  {author} {\bibfnamefont {G.}~\bibnamefont {Semeghini}}, \bibinfo {author}
  {\bibfnamefont {M.}~\bibnamefont {Greiner}}, \bibinfo {author} {\bibfnamefont
  {V.}~\bibnamefont {Vuletić}},\ and\ \bibinfo {author} {\bibfnamefont
  {M.~D.}\ \bibnamefont {Lukin}},\ }\bibfield  {title} {\bibinfo {title}
  {High-fidelity parallel entangling gates on a neutral-atom quantum
  computer},\ }\href {https://doi.org/10.1038/s41586-023-06481-y} {\bibfield
  {journal} {\bibinfo  {journal} {Nature}\ }\textbf {\bibinfo {volume} {622}},\
  \bibinfo {pages} {268–272} (\bibinfo {year} {2023})}\BibitemShut {NoStop}%
\bibitem [{\citenamefont {Gidney}\ \emph {et~al.}(2021)\citenamefont {Gidney},
  \citenamefont {Newman}, \citenamefont {Fowler},\ and\ \citenamefont
  {Broughton}}]{gidney2021b}%
  \BibitemOpen
  \bibfield  {author} {\bibinfo {author} {\bibfnamefont {C.}~\bibnamefont
  {Gidney}}, \bibinfo {author} {\bibfnamefont {M.}~\bibnamefont {Newman}},
  \bibinfo {author} {\bibfnamefont {A.}~\bibnamefont {Fowler}},\ and\ \bibinfo
  {author} {\bibfnamefont {M.}~\bibnamefont {Broughton}},\ }\bibfield  {title}
  {\bibinfo {title} {A {{Fault-Tolerant Honeycomb Memory}}},\ }\href
  {https://doi.org/10.22331/q-2021-12-20-605} {\bibfield  {journal} {\bibinfo
  {journal} {Quantum}\ }\textbf {\bibinfo {volume} {5}},\ \bibinfo {pages}
  {605} (\bibinfo {year} {2021})}\BibitemShut {NoStop}%
\bibitem [{\citenamefont {Xu}\ \emph {et~al.}(2023)\citenamefont {Xu},
  \citenamefont {Ataides}, \citenamefont {Pattison}, \citenamefont
  {Raveendran}, \citenamefont {Bluvstein}, \citenamefont {Wurtz}, \citenamefont
  {Vasic}, \citenamefont {Lukin}, \citenamefont {Jiang},\ and\ \citenamefont
  {Zhou}}]{xu2023}%
  \BibitemOpen
  \bibfield  {author} {\bibinfo {author} {\bibfnamefont {Q.}~\bibnamefont
  {Xu}}, \bibinfo {author} {\bibfnamefont {J.~P.~B.}\ \bibnamefont {Ataides}},
  \bibinfo {author} {\bibfnamefont {C.~A.}\ \bibnamefont {Pattison}}, \bibinfo
  {author} {\bibfnamefont {N.}~\bibnamefont {Raveendran}}, \bibinfo {author}
  {\bibfnamefont {D.}~\bibnamefont {Bluvstein}}, \bibinfo {author}
  {\bibfnamefont {J.}~\bibnamefont {Wurtz}}, \bibinfo {author} {\bibfnamefont
  {B.}~\bibnamefont {Vasic}}, \bibinfo {author} {\bibfnamefont {M.~D.}\
  \bibnamefont {Lukin}}, \bibinfo {author} {\bibfnamefont {L.}~\bibnamefont
  {Jiang}},\ and\ \bibinfo {author} {\bibfnamefont {H.}~\bibnamefont {Zhou}},\
  }\href {http://arxiv.org/abs/2308.08648} {\bibinfo {title}
  {Constant-{{Overhead Fault-Tolerant Quantum Computation}} with
  {{Reconfigurable Atom Arrays}}}} (\bibinfo {year} {2023}),\ \Eprint
  {https://arxiv.org/abs/2308.08648} {arxiv:2308.08648 [quant-ph]} \BibitemShut
  {NoStop}%
\bibitem [{\citenamefont {Horsman}\ \emph {et~al.}(2012)\citenamefont
  {Horsman}, \citenamefont {Fowler}, \citenamefont {Devitt},\ and\
  \citenamefont {Meter}}]{horsman2012}%
  \BibitemOpen
  \bibfield  {author} {\bibinfo {author} {\bibfnamefont {C.}~\bibnamefont
  {Horsman}}, \bibinfo {author} {\bibfnamefont {A.~G.}\ \bibnamefont {Fowler}},
  \bibinfo {author} {\bibfnamefont {S.}~\bibnamefont {Devitt}},\ and\ \bibinfo
  {author} {\bibfnamefont {R.~V.}\ \bibnamefont {Meter}},\ }\bibfield  {title}
  {\bibinfo {title} {Surface code quantum computing by lattice surgery},\
  }\href {https://doi.org/10.1088/1367-2630/14/12/123011} {\bibfield  {journal}
  {\bibinfo  {journal} {New Journal of Physics}\ }\textbf {\bibinfo {volume}
  {14}},\ \bibinfo {pages} {123011} (\bibinfo {year} {2012})}\BibitemShut
  {NoStop}%
\bibitem [{\citenamefont {Ramette}\ \emph {et~al.}(2023)\citenamefont
  {Ramette}, \citenamefont {Sinclair}, \citenamefont {Breuckmann},\ and\
  \citenamefont {Vuleti{\'c}}}]{ramette2023}%
  \BibitemOpen
  \bibfield  {author} {\bibinfo {author} {\bibfnamefont {J.}~\bibnamefont
  {Ramette}}, \bibinfo {author} {\bibfnamefont {J.}~\bibnamefont {Sinclair}},
  \bibinfo {author} {\bibfnamefont {N.~P.}\ \bibnamefont {Breuckmann}},\ and\
  \bibinfo {author} {\bibfnamefont {V.}~\bibnamefont {Vuleti{\'c}}},\ }\href
  {https://doi.org/10.48550/arXiv.2302.01296} {\bibinfo {title}
  {Fault-{{Tolerant Connection}} of {{Error-Corrected Qubits}} with {{Noisy
  Links}}}} (\bibinfo {year} {2023}),\ \Eprint
  {https://arxiv.org/abs/2302.01296} {arxiv:2302.01296 [math-ph,
  physics:quant-ph]} \BibitemShut {NoStop}%
\bibitem [{\citenamefont {Breuckmann}\ and\ \citenamefont
  {Eberhardt}(2021)}]{breuckmann2021}%
  \BibitemOpen
  \bibfield  {author} {\bibinfo {author} {\bibfnamefont {N.~P.}\ \bibnamefont
  {Breuckmann}}\ and\ \bibinfo {author} {\bibfnamefont {J.~N.}\ \bibnamefont
  {Eberhardt}},\ }\bibfield  {title} {\bibinfo {title} {Quantum {{Low-Density
  Parity-Check Codes}}},\ }\href@noop {} {\bibfield  {journal} {\bibinfo
  {journal} {PRX Quantum}\ }\textbf {\bibinfo {volume} {2}},\ \bibinfo {pages}
  {21} (\bibinfo {year} {2021})}\BibitemShut {NoStop}%
\bibitem [{\citenamefont {Bowers}\ \emph {et~al.}(1996)\citenamefont {Bowers},
  \citenamefont {Budker}, \citenamefont {Commins}, \citenamefont {DeMille},
  \citenamefont {Freedman}, \citenamefont {Nguyen}, \citenamefont {Shang},\
  and\ \citenamefont {Zolotorev}}]{bowers1996}%
  \BibitemOpen
  \bibfield  {author} {\bibinfo {author} {\bibfnamefont {C.~J.}\ \bibnamefont
  {Bowers}}, \bibinfo {author} {\bibfnamefont {D.}~\bibnamefont {Budker}},
  \bibinfo {author} {\bibfnamefont {E.~D.}\ \bibnamefont {Commins}}, \bibinfo
  {author} {\bibfnamefont {D.}~\bibnamefont {DeMille}}, \bibinfo {author}
  {\bibfnamefont {S.~J.}\ \bibnamefont {Freedman}}, \bibinfo {author}
  {\bibfnamefont {A.~T.}\ \bibnamefont {Nguyen}}, \bibinfo {author}
  {\bibfnamefont {S.~Q.}\ \bibnamefont {Shang}},\ and\ \bibinfo {author}
  {\bibfnamefont {M.}~\bibnamefont {Zolotorev}},\ }\bibfield  {title} {\bibinfo
  {title} {Experimental investigation of excited-state lifetimes in atomic
  ytterbium},\ }\href {http://link.aps.org/doi/10.1103/PhysRevA.53.3103}
  {\bibfield  {journal} {\bibinfo  {journal} {Physical Review A}\ }\textbf
  {\bibinfo {volume} {53}},\ \bibinfo {pages} {3103} (\bibinfo {year}
  {1996})}\BibitemShut {NoStop}%
\bibitem [{\citenamefont {Jaffe}\ \emph {et~al.}(2022)\citenamefont {Jaffe},
  \citenamefont {Palm}, \citenamefont {Baum}, \citenamefont {Taneja},
  \citenamefont {Kumar},\ and\ \citenamefont {Simon}}]{jaffe2022a}%
  \BibitemOpen
  \bibfield  {author} {\bibinfo {author} {\bibfnamefont {M.}~\bibnamefont
  {Jaffe}}, \bibinfo {author} {\bibfnamefont {L.}~\bibnamefont {Palm}},
  \bibinfo {author} {\bibfnamefont {C.}~\bibnamefont {Baum}}, \bibinfo {author}
  {\bibfnamefont {L.}~\bibnamefont {Taneja}}, \bibinfo {author} {\bibfnamefont
  {A.}~\bibnamefont {Kumar}},\ and\ \bibinfo {author} {\bibfnamefont
  {J.}~\bibnamefont {Simon}},\ }\bibfield  {title} {\bibinfo {title}
  {Understanding and suppressing backscatter in optical resonators},\ }\href
  {https://doi.org/10.1364/OPTICA.463723} {\bibfield  {journal} {\bibinfo
  {journal} {Optica}\ }\textbf {\bibinfo {volume} {9}},\ \bibinfo {pages} {878}
  (\bibinfo {year} {2022})}\BibitemShut {NoStop}%
\bibitem [{\citenamefont {{Tanji-Suzuki}}\ \emph {et~al.}(2011)\citenamefont
  {{Tanji-Suzuki}}, \citenamefont {Leroux}, \citenamefont {{Schleier-Smith}},
  \citenamefont {Cetina}, \citenamefont {Grier}, \citenamefont {Simon},\ and\
  \citenamefont {Vuleti{\'c}}}]{tanji-suzuki2011}%
  \BibitemOpen
  \bibfield  {author} {\bibinfo {author} {\bibfnamefont {H.}~\bibnamefont
  {{Tanji-Suzuki}}}, \bibinfo {author} {\bibfnamefont {I.~D.}\ \bibnamefont
  {Leroux}}, \bibinfo {author} {\bibfnamefont {M.~H.}\ \bibnamefont
  {{Schleier-Smith}}}, \bibinfo {author} {\bibfnamefont {M.}~\bibnamefont
  {Cetina}}, \bibinfo {author} {\bibfnamefont {A.~T.}\ \bibnamefont {Grier}},
  \bibinfo {author} {\bibfnamefont {J.}~\bibnamefont {Simon}},\ and\ \bibinfo
  {author} {\bibfnamefont {V.}~\bibnamefont {Vuleti{\'c}}},\ }\bibfield
  {title} {\bibinfo {title} {Interaction between {{Atomic Ensembles}} and
  {{Optical Resonators}}},\ }in\ \href
  {https://doi.org/10.1016/B978-0-12-385508-4.00004-8} {\emph {\bibinfo
  {booktitle} {Advances {{In Atomic}}, {{Molecular}}, and {{Optical
  Physics}}}}},\ Vol.~\bibinfo {volume} {60}\ (\bibinfo  {publisher}
  {{Elsevier}},\ \bibinfo {year} {2011})\ pp.\ \bibinfo {pages}
  {201--237}\BibitemShut {NoStop}%
\bibitem [{\citenamefont {Gardiner}\ and\ \citenamefont
  {Zoller}(2004)}]{gardiner2004quantum}%
  \BibitemOpen
  \bibfield  {author} {\bibinfo {author} {\bibfnamefont {C.}~\bibnamefont
  {Gardiner}}\ and\ \bibinfo {author} {\bibfnamefont {P.}~\bibnamefont
  {Zoller}},\ }\href@noop {} {\emph {\bibinfo {title} {Quantum noise: a
  handbook of Markovian and non-Markovian quantum stochastic methods with
  applications to quantum optics}}}\ (\bibinfo  {publisher} {Springer Science
  \& Business Media},\ \bibinfo {year} {2004})\BibitemShut {NoStop}%
\bibitem [{\citenamefont {Kiraz}\ \emph {et~al.}(2004)\citenamefont {Kiraz},
  \citenamefont {Atatüre},\ and\ \citenamefont {Imamoğlu}}]{Kiraz2004}%
  \BibitemOpen
  \bibfield  {author} {\bibinfo {author} {\bibfnamefont {A.}~\bibnamefont
  {Kiraz}}, \bibinfo {author} {\bibfnamefont {M.}~\bibnamefont {Atatüre}},\
  and\ \bibinfo {author} {\bibfnamefont {A.}~\bibnamefont {Imamoğlu}},\
  }\bibfield  {title} {\bibinfo {title} {Quantum-dot single-photon sources:
  Prospects for applications in linear optics quantum-information processing},\
  }\href {https://doi.org/10.1103/PhysRevA.69.032305} {\bibfield  {journal}
  {\bibinfo  {journal} {Physical Review A}\ }\textbf {\bibinfo {volume} {69}},\
  \bibinfo {pages} {032305} (\bibinfo {year} {2004})}\BibitemShut {NoStop}%
\bibitem [{\citenamefont {Johansson}\ \emph {et~al.}(2013)\citenamefont
  {Johansson}, \citenamefont {Nation},\ and\ \citenamefont {Nori}}]{qutip}%
  \BibitemOpen
  \bibfield  {author} {\bibinfo {author} {\bibfnamefont {J.}~\bibnamefont
  {Johansson}}, \bibinfo {author} {\bibfnamefont {P.}~\bibnamefont {Nation}},\
  and\ \bibinfo {author} {\bibfnamefont {F.}~\bibnamefont {Nori}},\ }\bibfield
  {title} {\bibinfo {title} {Qutip 2: A python framework for the dynamics of
  open quantum systems},\ }\href
  {https://doi.org/https://doi.org/10.1016/j.cpc.2012.11.019} {\bibfield
  {journal} {\bibinfo  {journal} {Computer Physics Communications}\ }\textbf
  {\bibinfo {volume} {184}},\ \bibinfo {pages} {1234} (\bibinfo {year}
  {2013})}\BibitemShut {NoStop}%
\end{thebibliography}%

\end{document}